%% file: FSGS_Scalar.tex
\documentclass[preprint]{elsarticle}

\usepackage[left=2.3cm, right=2.3cm, top=2.5cm, bottom=2.5cm]{geometry}

\graphicspath{{figures/}}
\input{preamble.tex}
\journal{Journal of Computational Physics}

\begin{document}

\begin{frontmatter}

\title{Data-Driven Fractional Subgrid-scale Modeling for Scalar Turbulence: \\ A Nonlocal LES Approach}

\author[ME,CMSE]{Ali Akhavan-Safaei}
\author[ME,CMSE]{Mehdi Samiee}
\author[ME,STAT]{Mohsen Zayernouri\corref{corr_author}}

\cortext[corr_author]{Corresponding author}\ead{zayern@msu.edu}

\address[ME]{Department of Mechanical Engineering, Michigan State University, East Lansing, MI 48824, USA}
\address[CMSE]{Department of Computational Mathematics, Science and Engineering, Michigan State University, East Lansing, MI 48824, USA}
\address[STAT]{Department of Statistics and Probability, Michigan State University, East Lansing, MI 48824, USA}

\begin{abstract}
Filtering the passive scalar transport equation in the large-eddy simulation (LES) of turbulent transport gives rise to the closure term corresponding to the unresolved scalar flux. Understanding and respecting the statistical features of subgrid-scale (SGS) flux is a crucial point in robustness and predictability of the LES. In this work, we investigate the intrinsic nonlocal behavior of the SGS passive scalar flux through studying its two-point statistics obtained from the filtered direct numerical simulation (DNS) data for passive scalar transport in homogeneous isotropic turbulence (HIT). Presence of long-range correlations in true SGS scalar flux urges to go beyond the conventional local closure modeling approaches that fail to predict the non-Gaussian statistical features of turbulent transport in passive scalars. Here, we propose an appropriate statistical model for microscopic SGS motions by taking into account the filtered Boltzmann transport equation (FBTE) for passive scalar. In FBTE, we approximate the filtered equilibrium distribution with an $\alpha$-stable L\'{e}vy distribution that essentially incorporates a power-law behavior to resemble the observed nonlocal statistics of SGS scalar flux. Generic ensemble-averaging of such FBTE lets us formulate a continuum level closure model for the SGS scalar flux appearing in terms of fractional-order Laplacian that is inherently nonlocal. Through a data-driven approach, we infer the optimal version of our SGS model using the high-fidelity data for the two-point correlation function between the SGS scalar flux and filtered scalar gradient, and sparse linear regression. In an \textit{a priori} test, the optimal fractional-order model yields a promising performance in reproducing the probability distribution function (PDF) of the SGS dissipation of the filtered scalar variance compared to its true PDF obtained from the filtered DNS data.
\end{abstract}


\end{frontmatter}


\section{Introduction}\label{sec: Intro}

Large-scale natural flows such as atmospheric ones, as well as a wide variety of engineering applications, are among many systems that are substantially influenced by turbulence. Nonlinearity and stochastisity are two inherent elements of fluid dynamics that when significantly triggered lead the flow into turbulent regime \citep{pope2001turbulent, sagaut2008homogeneous}. Turbulence is characterized by persistent fluctuating field variables that are immensely non-Gaussian and have a multi-scale and ubiquitous influence on the fluid dynamics with a great impact on the quality of transport and mixing \citep{akhavan2020anomalous, zayernouri2011coherent}. Moreover, notable emergence of the extreme and anomalous events reflected in the statistical measurements of turbulent fields and quantities intensify the level of complexity in turbulent flows \citep{sapsis2020statistics, yeung2015extreme}. Therefore, taking into account the effects of turbulence cannot be compromised in predictions and design procedure for a fluid system affected by the turbulent regime. Although considerable advancements in the modern computational architectures and high-performance computing (HPC) over the past decade have greatly facilitated the high-fidelity predictions of turbulent transport through direct numerical simulation (DNS), those efforts have  mostly remained in the area of canonical and fundamental turbulent transport. Nevertheless, large-eddy simulation (LES) of turbulence has shown a promising path towards robust, accurate, and computationally affordable predictions of the turbulent flow behavior in large-scale and real-world applications \citep{fu2020heat}. In fact, LES is considered as a reliable trade-off between the DNS and the low-fidelity simulations with Reynolds-Averaged Navier-Stokes (RANS) models. The main idea in the LES is that for sufficiently high-Reynolds flows that the statistics of turbulent fluctuations associated with small-scale motions are isotropic and hence we expect a universal behavior, one can numerically resolve the large-scale motions while dealing with the subgrid-scale (SGS) effects through proper closure modeling means that utilize resolved-scale variables. In practice, there is a spatial filtering acting on the conservation equations of transport that represents the LES equations \citep{leonard1975energy, germano1992turbulence}. 

Traditionally, SGS modeling is categorized into two main branches: (\textit{i}) functional modeling, and (\textit{ii}) structural modeling \citep{sagaut2006large}. Functional modeling requires a prior knowledge of the interactions between resolved-scale and subgrid-scale is required so that one can represent the LES closure in terms of a mathematical function of resolved transport variables. Functional models are usually representing the net transfer of turbulent kinetic energy from resolved scales to the subgrid scales. The Smagorinsky model initially conceptualized in \citep{smagorinsky1963general}, and its variations are well-known examples of functional SGS modeling. On the other hand, structural models seek to reconstruct the statics and structure of SGS stresses and fluxes from the resolved-scale variables. For instance, scale-similarity models initially introduced by Bardina \textit{et al.} \citep{bardina1980improved} are among well-known examples of structural models. Functional models usually are poorly correlated with the true SGS terms \textit{a priori} and by construction are incapable of reproducing backward transfer of energy (backscattering); however, in an LES setting they have shown to be dissipative enough for solver stability. In contrast, structural models such as scale-similarity type models have been found to be sufficiently correlated with the true SGS terms and fairly capable of following backscattering phenomenon in an \textit{a priori} sense. Nonetheless, their significant drawback is that in LES they are under-dissipative; hence, the stable time-integration is intractable. As a practical remedy to the mentioned issues, further efforts have been devoted to formulating a mixed representation of functional and structural models \citep{zang1993dynamic, liu1994properties}. Recently, abundance of the high-fidelity data for the SGS closures mainly available through filtered DNS data, and with the advent of modern machine learning (ML) techniques and their application to fluid mechanics and in particular, turbulence modeling, \citep{kutz2017deep, duraisamy2019turbulence, brunton2020machine, beck2020perspective} have resulted in a wide variety of predictive data-driven SGS models. Among the numerous contributions in ML-based SGS modeling and LES, interested readers are referred to the following notable works \citep{beck2019deep, kurz2020machine,portwood2020interpreting, SIRIGNANO2020}.  

A vital point to credibly certify an SGS model for the LES is its capability to accurately encode the statistics of turbulent transport and SGS dynamics \citep{meneveau1994statistics, moser2020statistical}. Therefore, in the current work, our main focus is to develop a statistically consistent LES closure model. Throughout this approach, we aim to ensures capturing the nonlocal interactions in the turbulent energy dissipation \citep{waleffe1992nature, hamlington2008local}, that are intensified in the SGS effects during LES \citep{samiee2020fractional}. 

Unlike the integer-order (standard) differential operators, fractional-order operators are fundamentally defined based on heavy-tailed stochastic processes; therefore, they are inherently nonlocal operators and are suitable to incorporate long-range interactions in a mathematical modeling \citep{meerschaert2011stochastic, d2020unified}. Among the wide-range of applications employing the fractional-order operators, modeling of visco-elasto-plastic materials for structural analysis \citep{suzuki2016fractional}, their nonlinear vibration analysis \citep{suzuki2020anomalous}, memory-dependent modeling of damage mechanics \citep{suzuki2019thermodynamically}, and nonlocal elasticity modeling of solids \citep{jokar2020variable} are listed among the outstanding works reported in the literature. Due to the remarkable and diverse applications of the fractional-order Partial Differential Equations (PDEs), development of high-order numerical methods \citep{samiee2017fast, samiee2019I, samiee2019II, lischke2019spectral, samiee2020unified, delia_nonlocal_2020, zhou2020implicit, du2020fast, fang2020fast} and data-driven numerical schemes \citep{suzuki2020self}, as well as numerical studies on the stochastic fractional PDEs \citep{kharazmi2019fractional, kharazmi2019operator} have been an active area of research.

In the context of LES for turbulent flows, a recent study by Samiee \textit{et al.} \citep{samiee2020fractional} introduced a nonlocal model for the divergence of SGS stress tensor in terms of a fractional Laplacian acting on the resolved-scale velocity field. In order to derive such a model, filtered Boltzmann transport equation was considered, where the filtered equilibrium distribution is approximated with an $\alpha$-stable L\'evy distribution. Moreover, Di Leoni \textit{et al.} \citep{di2020two} proposed a nonlocal eddy-viscosity SGS model that employs a fractional gradient operator. Their modeling strategy is based upon the high-fidelity observations of nonlocal two-point correlation between the SGS stress and strain-rate tensors (inspired by the derivation of filtered K\'arman-Howarth equation), and proposing a proper nonlocal convolution kernel that yields the fractional gradient operator. They sufficiently captured the nonlocal SGS effects through proper fractional orders for different turbulent flows including the anisotropy and inhomogeneity effects. These studies demonstrated that the fractional-order operators are sophisticated candidates for modeling of the SGS stresses in the LES of turbulent flows. 

Of particular interest, we aim to study the nonlocal SGS modeling for the conserved passive scalars in turbulent flows \citep{warhaft2000passive, shraiman2000scalar, sreenivasan2019turbulent}; thus, we seek to model the SGS scalar flux arising as the closure term in the filtered scalar transport equation. Due to promising potential of Boltzmann transport framework to investigate the sources of spatial nonlocality appearing in the SGS dynamics \citep{samiee2020fractional}, we manage to study the filtered version of the Boltzmann transport equation for the passive scalars in turbulent flow. Using proper statistical assumptions at the kinetic level, we try to derive a continuum level closure model in terms of fractional-order Laplacian of the resolved scalar concentration. Through a statistical data-driven procedure our model is being calibrated to its optimal form so that it is capable of capturing the nonlocal statistics embedded in the ground-truth data.


The structure of the rest of this work is organized as follows: in section \ref{sec: Gov-Eqns}, we state the problem and show the governing equations. In section \ref{sec: nonlocality}, we motivate the necessity of our modeling strategy to address nonlocality using statistical measures obtained from the filtered DNS data. In section \ref{sec: BT-framework}, the mathematical framework of our SGS modeling that includes fractional calculus and Boltzmann transport is described and derivation of the SGS model is presented. Afterwards, in section \ref{sec: Calibration}, a two-stage data-driven calibration procedure is introduced to optimize the model performance. Finally, section \ref{sec: Dissipation} delivers an \textit{a priori} testing on the SGS dissipation of the resolved-scale scalar variance, and it is followed by the conclusions in section \ref{sec: Conclusion}.

\section{Governing Equations}\label{sec: Gov-Eqns}

Considering flows governed by incompressible Navier-Stokes (NS) equations
\begin{eqnarray}\label{GE-1-2}
    \frac{\partial V_i}{\partial t}+\frac{\partial}{\partial x_j}\left( V_i\,V_j \right)=-\frac{1}{\rho}\frac{\partial p}{\partial x_i}+\nu \, \frac{\partial^2 V_i}{\partial x_i \partial x_j}+\mathcal{A} \, V_i, \quad i,j=1,2,3,
\end{eqnarray}
subject to the continuity, $\nabla \cdot \boldsymbol{V}=0$, where the velocity and the pressure fields are denoted by $\boldsymbol{V}(\boldsymbol{x},t)=(V_1,\, V_2,\, V_3)$ and $p(\boldsymbol{x},t)$ for $\boldsymbol{x}=x_i$ and $i=1,2,3$, respectively. $\rho$ specifies the density and $\nu$ represents the kinematic viscosity for a Newtonian fluid. In \eqref{GE-1-2}, $\mathcal{A}$ is a dynamic coefficient associated with the artificial forcing scheme to enforce statistical stationary state on the kinetic energy to reach to a realistic and fully turbulent state. It is worth mentioning that all the values in \eqref{GE-1-2} are taken to be zero-mean values, therefore, $\boldsymbol{V}(\boldsymbol{x},t)$ corresponds to the turbulent fluctuations. In our study, a passive scalar with an imposed mean gradient along the $x_2$ direction is considered to be transported with the described turbulent flow. According to the Reynolds decomposition for the total concentration of the passive scalar, $\Phi(\boldsymbol{x},t)$, one can write that $\Phi = \langle \Phi \rangle + \phi$. Here, $\langle \cdot \rangle$ is the ensemble-averaging operator, and $\phi$ denotes the fluctuating part of the passive scalar concentration. More specifically, the imposed mean scalar gradient is taken to be uniform as $\nabla  \langle \Phi \rangle = \left( 0,\beta,0\right)$, where $\beta$ is a constant. Therefore, the turbulent scalar concentration obeys an advection-diffusion (AD) equation that is simplified into the following form
\begin{eqnarray}\label{GE-1-3}
    \frac{\partial \phi}{\partial t}+\frac{\partial}{\partial x_i}\left( \phi \, V_i \right) = -\beta \, V_2+\mathcal{D} \, \frac{\partial^2 \phi}{\partial x_i \partial x_i}, \quad i=1,2,3,
\end{eqnarray}
where $\mathcal{D}$ denotes the molecular diffusion coefficient of the passive scalar. Accordingly, the Schmidt number is defined as $Sc=\nu/\mathcal{D}$.

In the LES of turbulent transport, the fluid and passive scalar motions are resolved down to a prescribed length scale namely as filter width,  $\Delta$,  which linearly decomposes the velocity and scalar concentration fields into the filtered (resolved) and the residual (unresolved) components. For instance, for the scalar concentration, $\widetilde{\phi}$ and $\phi^R=\phi - \widetilde{\phi}$ represent the filtered and residual fields, respectively. The filtered fields are obtained by a convolution, $\widetilde{\phi}= \boldsymbol{\mathcal{G}} \ast \phi$, where $\boldsymbol{\mathcal{G}} = \boldsymbol{\mathcal{G}}(\boldsymbol{r})$ denotes the generic spatial filtering kernel \citep{pope2001turbulent}. Applying such filtering operation on the governing equations returns the subsequent LES equations. For example, the filtered AD equation is formulated as 
\begin{eqnarray}\label{eqn: Filtered-AD}
    \frac{\partial \widetilde{\phi}}{\partial t}+\frac{\partial}{\partial x_i}\left( \widetilde{\phi}\,\widetilde{V}_i \right) = -\beta \, \widetilde{V}_2 + \mathcal{D} \, \frac{\partial^2 \widetilde{\phi}}{\partial x_i \partial x_i} - \frac{\partial q^R_i}{\partial x_i}, \quad i=1,2,3,
\end{eqnarray}
where $q^R_i$ denotes the residual or SGS scalar flux that is defined exactly as $q^R_i=\widetilde{\phi \, V_i}-\widetilde{\phi} \, \widetilde{V}_i$. In the LES sense, the SGS scalar flux needs to be closed (modeled) in terms of the resolved-scale (filtered) variables through proper and physically consistent SGS modeling.

\section{Why SGS Dynamics is Statistically Nonlocal?}\label{sec: nonlocality}

\begin{figure}[t!]
    \begin{minipage}[b]{.48\linewidth}
        \centering
        \includegraphics[width=1\textwidth]{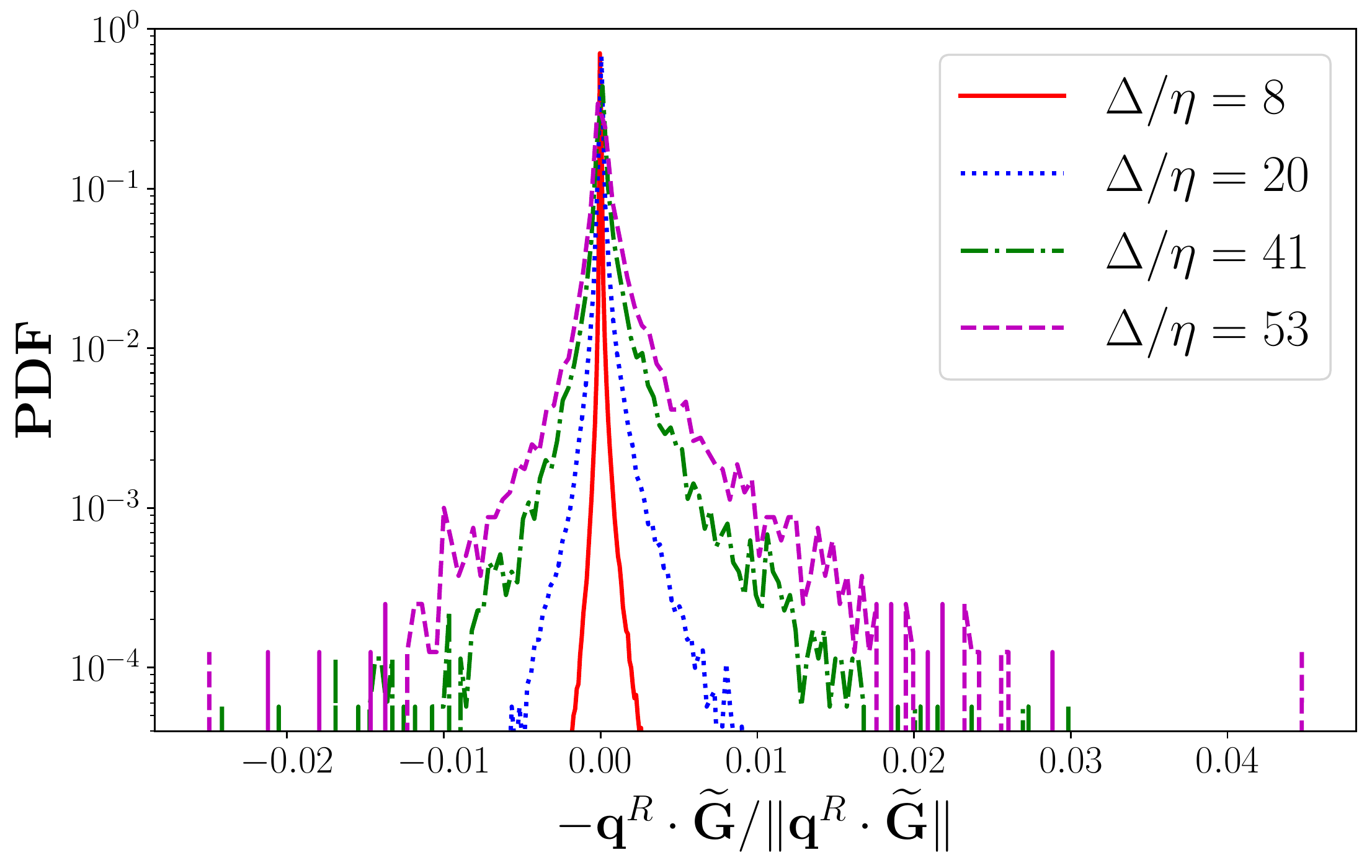}
        \subcaption{\footnotesize}
    \end{minipage}
    \begin{minipage}[b]{.01\linewidth}
    		~
     \end{minipage}
    \begin{minipage}[b]{.49\linewidth}
        \centering
        \includegraphics[width=1\textwidth]{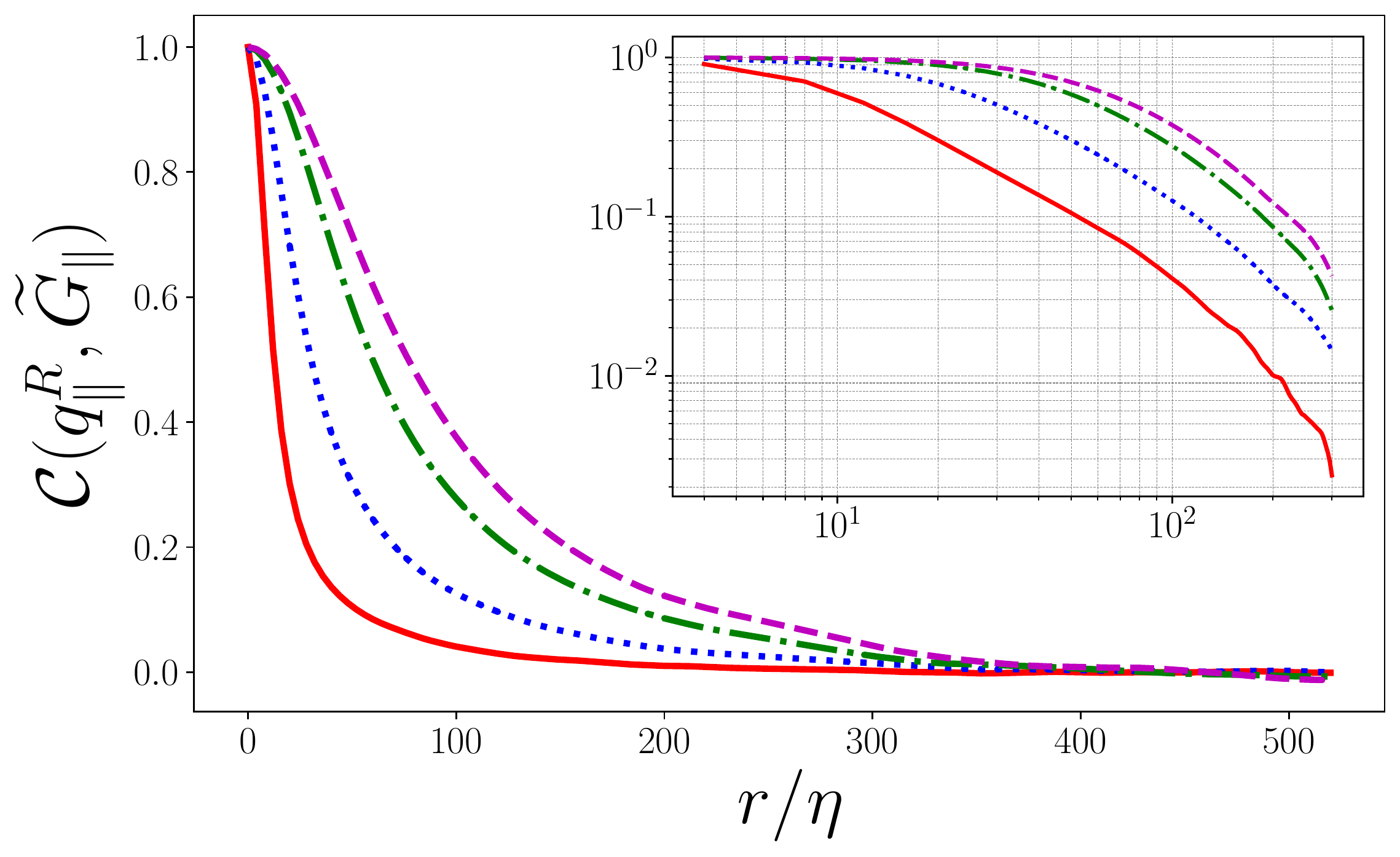}
        \subcaption{\footnotesize}
    \end{minipage}   
    \caption{\footnotesize Statistics of true subgrid-scale contribution to the filtered scalar variance rate. (a) PDF of normalized SGS dissipation of filtered scalar variance, $-\boldsymbol{q}^R \cdot \widetilde{\boldsymbol{G}}$, computed over a sample space of $10 \, T_{LE}$ of statically stationary turbulence. (b) Time-averaged two-point correlation function \eqref{eqn: TPC1} between $q^R_\parallel$ and $\widetilde{G}_\parallel$ with $r=\vert \boldsymbol{r}_\perp \vert$.}\label{fig: Nonlocality}
\end{figure}

In an idealistic LES, one of the main elements reflecting the dynamics of turbulent transport is capturing the true filtered (resolved-scale) turbulent intensity through robust SGS modeling that is physically and mathematically consistent. In fact, such transport equation includes closure terms that directly link the correct time-evolution of turbulent intensity to the nature of the SGS closure and its modeling. In the LES of scalar turbulence, multiplying both sides of the filtered AD equation \eqref{eqn: Filtered-AD} by $\widetilde{\phi}$, yields the time evolution of filtered turbulent \textit{intensity} as
\begin{equation}\label{eqn: scalar_var1}
	\frac{1}{2} \frac{\partial}{\partial t}\left( \widetilde{\phi} \, \widetilde{\phi} \right) + \widetilde{\phi} \, \frac{\partial}{\partial x_i} \left( \widetilde{\phi} \, \widetilde{V}_i \right)= -\beta \, \widetilde{\phi} \, \widetilde{V}_2 + \mathcal{D} \, \widetilde{\phi} \, \frac{\partial^2 \, \widetilde{\phi}}{\partial x_i \partial x_i} - \widetilde{\phi} \, \frac{\partial \, q^R_i}{\partial x_i}.
\end{equation}
Using the continuity equation and chain rule for differentiation,
\begin{equation}\label{eqn: scalar_var2}
	\frac{1}{2} \frac{\partial}{\partial t}\left( \widetilde{\phi} \, \widetilde{\phi} \right) + \widetilde{\phi} \, \widetilde{V}_i \, \frac{\partial \widetilde{\phi}}{\partial x_i} = -\beta \, \widetilde{\phi} \, \widetilde{V}_2 + \mathcal{D} \, \frac{\partial}{\partial x_i}\left( \widetilde{\phi} \, \frac{\partial \widetilde{\phi}}{\partial x_i} \right) - \mathcal{D} \, \frac{\partial \, \widetilde{\phi}}{\partial x_i} \, \frac{\partial \, \widetilde{\phi}}{\partial x_i} - \frac{\partial}{\partial x_i}\left( \widetilde{\phi} \, q^R_i \right) + q^R_i \, \frac{\partial \widetilde{\phi}}{\partial x_i}.
\end{equation}
Applying the ensemble-averaging operator, $\langle \cdot \rangle$, on \eqref{eqn: scalar_var2}, returns the transport equation for the \textit{filtered scalar variance}, $\left\langle \widetilde{\phi} \, \widetilde{\phi} \right\rangle$. In this study, we are considering the case of homogeneous turbulent velocity and scalar fields; therefore, $\left\langle \frac{\partial}{\partial x_i}\left(\cdot \right) \right\rangle = \frac{\partial}{\partial x_i}\langle (\cdot) \rangle = 0$. Defining the filtered scalar gradient as $\widetilde{\boldsymbol{G}}(\boldsymbol{x}) = \nabla \widetilde{\phi}(\boldsymbol{x})$, time-evolution of the filtered scalar variance takes the following form 
\begin{align}\label{eqn: scalar_var3}
    	\frac{1}{2} \frac{d}{d t}\left\langle \widetilde{\phi} \, \widetilde{\phi} \right\rangle &= -\widetilde{\mathcal{T}} + \widetilde{\mathcal{P}} - \widetilde{\chi} + \Pi, \\
    	\widetilde{\mathcal{T}} = \left\langle \widetilde{\phi} \, \widetilde{V}_i \, \widetilde{G}_i \right \rangle, \quad \widetilde{\mathcal{P}} = -\beta \left\langle \widetilde{\phi} \, \widetilde{V}_2 \right \rangle&, \quad \widetilde{\chi} = \mathcal{D} \, \left\langle \widetilde{G}_i \, \widetilde{G}_i \right\rangle, \quad \Pi = \left\langle q^R_i \, \widetilde{G}_i \right\rangle. \nonumber
\end{align}
In \eqref{eqn: scalar_var3}, $\widetilde{\mathcal{T}}$ denotes the \text{turbulent transport} of filtered scalar variance while $\widetilde{\mathcal{P}}$ represents the \textit{production} of resolved scalar variance by the uniform mean scalar gradient, and $\widetilde{\chi}$ is the resolved scalar variance \text{dissipation} due to the molecular diffusion. Unlike these three terms, $\Pi$ (representing the \textit{SGS production} of resolved scalar variance) is the only contributing term in \eqref{eqn: scalar_var3} that contains the effects of the SGS scalar flux. Therefore, as pointed out earlier, understanding the true statistical nature of $\boldsymbol{q}^R \cdot \widetilde{\boldsymbol{G}}$ is essential for the SGS modeling and precise evaluation of the resolved scalar variance in the LES. This examination of $\boldsymbol{q}^R \cdot \widetilde{\boldsymbol{G}}$ might be viewed both from single-point and two-point statistics as discussed in \citep{meneveau1994statistics} in the context of the LES for homogeneous isotropic turbulent flows. In a recent comprehensive study by Di Leoni \textit{et al.}, effects of the SGS contribution in the evolution of the two-point velocity correlation was explored for the incompressible Navier-Stokes equations using filtered DNS data for HIT and turbulent channel flows at high-Reynolds numbers, and revealed the importance of nonlocal effects in the SGS dynamics \citep{di2020two}. In the present study, we are also focused on the two-point statistics of the SGS production of resolved scalar variance. This quantity is well represented in terms of the following normalized two-point correlation function 
\begin{align}\label{eqn: TPC1}
    \mathcal{C}(q^R_i \, , \, \widetilde{G}_i) = \frac{\left \langle q^R_i(\boldsymbol{x}) \, \widetilde{G}_i(\boldsymbol{x}+\boldsymbol{r}) \right \rangle}{\left \langle q^R_i(\boldsymbol{x}) \, \widetilde{G}_i(\boldsymbol{x}) \right \rangle},
\end{align}
where $\boldsymbol{r}=(r_1,r_2,r_3)$ denotes the spatial shift from the location $\boldsymbol{x}$. Moreover, probability distribution function (PDF) of the SGS production of scalar variance normalized by its $L_2$-norm \textit{i.e.}, $\boldsymbol{q}^R \cdot \widetilde{\boldsymbol{G}}/\Vert \boldsymbol{q}^R \cdot \widetilde{\boldsymbol{G}} \Vert$, is another measure to learn about the statistical behavior of $\Pi$ and have a more comprehensive insight into the SGS modeling. 

\subsection{High-Fidelity Database of the SGS Scalar Flux}\label{sec: Filtered-DNS}
In order to study the statistics of $\Pi$, we compute true values of the SGS scalar flux using the box filtering kernel with isotropic filter width $\Delta$ as,
\begin{align}\label{eqn: Box-filtering}
    \mathcal{G}(r) = 
    \begin{cases}
        \frac{1}{\Delta}, & r \leq \Delta/2 \\
        0, & r \geq \Delta/2.
    \end{cases}
\end{align}
Applying this convolution kernel on a well-resolved DNS database of passive scalar with imposed mean gradient in synthetic (forced) homogeneous isotropic turbulence. To perform the simulation, we employ an open-source parallel statistical-computational platform for turbulent transport equipped with a Fourier pseudo-spectral spatial discretization of the NS and AD equations, fourth-order Runge-Kutta (RK4) time-integration scheme, and an artificial forcing method (to keep the turbulent kinetic energy at low wavenumbers constant) \citep{PSc_HIT3D2020}. Our computational domain is a triply periodic cube of $\boldsymbol{\Omega}=[0,2\pi]^3$ that is discretized on a uniform Cartesian grid with $N=520^3$ Fourier collocation points while a constant $\Delta t = 5 \times 10^{-4}$ is utilized for the stable time-integration. In construction of this DNS database, the imposed mean scalar gradient is taken as $\beta=1$, and $Sc=1$ according to the section \ref{sec: Gov-Eqns}. Letting $k_{max}$ be the maximum resolved wavenumber in our simulation and $\eta=(\nu^3/\varepsilon)^{1/4}$ be the Kolmogorov length scale while $\varepsilon$ denotes the turbulent dissipation rate, we measure $k_{max} \, \eta \approx 1.5$; therefore, one can ensure that the small-scales in the velocity and scalar fields are well-resolved \citep{PSc_HIT3D2020}. Moreover, our records indicate that the Taylor-scale Reynolds number is $Re_\lambda=240$ (averaged over 25 large-eddy turnover times, $T_{LE}$, of resolving the passive scalar field).

\subsection{Statistical Analysis of the SGS Effects in Filtered Scalar Intensity}\label{sec: Stats-FDNS}
By taking a large sample space over $10 \, T_{LE}$ of this stationary process (after resolving the passive scalar field for $15 \, T_{LE}$), we compute the PDF of the normalized SGS production of filtered scalar variance for four different filter widths, $\Delta/\eta=8, \, 20, \, 41, \, 53$. As a result, we observe that as $\Delta$ becomes larger the PDF exhibits broader tails as shown in Figure \ref{fig: Nonlocality}(a). Emergence of these heavy PDF tails implies that as we increase the filter width, long-range spatial interactions become stronger and more pronounced \citep{akhavan2020anomalous}. Motivated by this observation, a two-point diagnosis of the SGS scalar production of the filtered variance as defined in equation \eqref{eqn: TPC1} would be another statistical measure shedding light on the long-range interactions in addition to the filter width effects. Considering $\parallel$ as the direction along the imposed mean scalar gradient and $\perp$ representing the directions perpendicular to the imposed mean gradient, we are interested in evaluating $\mathcal{C}(q^R_\parallel \, , \, \widetilde{G}_\parallel)$. Here, we take $\boldsymbol{r}=(r_1,0,0)$ and $\boldsymbol{r}=(0,0,r_3)$ and take the average of the resulting two-point correlation functions. Due to the statistically stationary turbulence, we perform such procedure for 20 data snapshots that are uniformly spaced over $10 \, T_{LE}$ (on the same spatio-temporal data we used to compute the PDFs); hence, we obtain the time-averaged value of $\mathcal{C}(q^R_\parallel \, , \, \widetilde{G}_\parallel)$. Figure \ref{fig: Nonlocality}(b) illustrates this two-point correlation function extending over a wide range of spatial shift, $r=\vert \boldsymbol{r} \vert$, and evaluated at four filter widths similar to the ones utilized in Figure \ref{fig: Nonlocality}(a). This plot quantitatively and qualitatively reveals that as we increase $\Delta$, greater correlation values between the SGS scalar flux $q^R_\parallel(\boldsymbol{x})$, and filtered scalar gradient $\widetilde{G}_\parallel(\boldsymbol{x}+\boldsymbol{r})$ are observed at a fixed $r$. These spatial correlations are significant both in the \textit{dissipation} and also \textit{inertial} subranges. This confirms the substantial nonlocal effects in the true SGS dynamics, which needs to be carefully addressed in the SGS modeling for LES. 

\begin{figure}[t!]
    \centering
    \includegraphics[width=.5\textwidth]{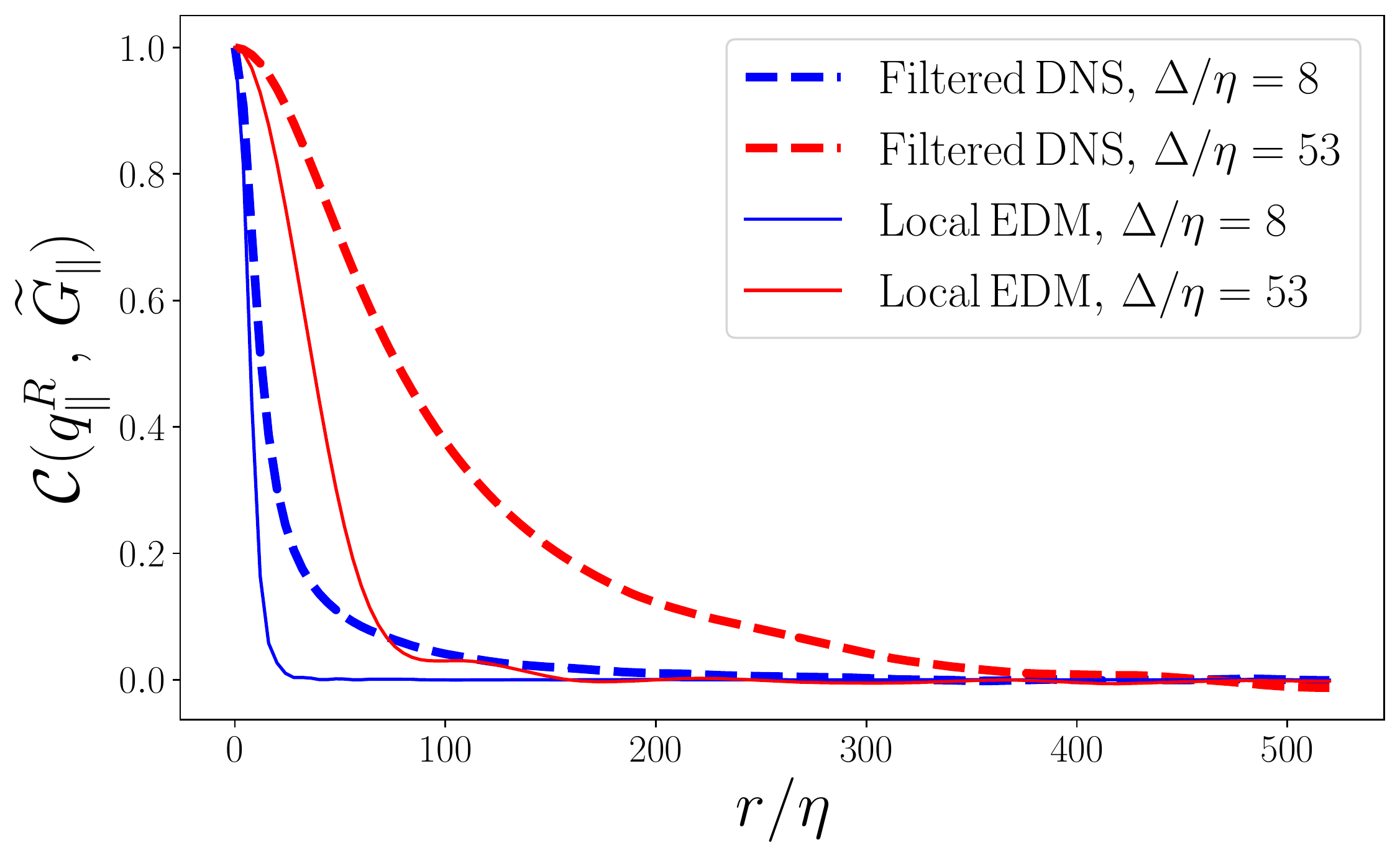}
    \caption{\footnotesize Comparison between the true values of two-point correlation function given in \eqref{eqn: TPC1} and the ones obtained from the local eddy-diffusivity modeling of the SGS scalar flux given in \eqref{eqn: EDM}. The evaluations are performed at two filter widths of $\Delta/\eta = 8, \, 53$.}\label{fig: EDM_TPC}
\end{figure}

A popular and fairly simple approach for modeling the SGS scalar flux is Eddy-Diffusivity Modeling (EDM). In EDM, the main assumption is that the SGS scalar flux is proportional to the resolved scale scalar gradient as
\begin{align}\label{eqn: EDM}
    \boldsymbol{q}^R(\boldsymbol{x}) \approx -\mathcal{D}_{ED} \, \widetilde{\boldsymbol{G}}(\boldsymbol{x}),  
\end{align}
and $\mathcal{D}_{ED}$ is the proportionality coefficient. Obviously, EDM is a \textit{local} modeling approach by its construction. Computing $\mathcal{C}(q^R_\parallel \, , \, \widetilde{G}_\parallel)$ while $q^R_\parallel$ is approximated with EDM, one can compare it with its true value as shown in Figure \ref{fig: Nonlocality}(b). Figure \ref{fig: EDM_TPC} illustrates such comparison for two filter widths, $\Delta/\eta=8, \, 53$, reveals that  
in both of the cases local EDM substantially fails to predict the conspicuous long-range spatial correlations observed in the true two-point correlation values. This observation is closely similar to the results reported in Di Leoni \textit{et al.} \citep{di2020two} that showed local eddy-viscosity model is structurally incapable of reproducing the two-point SGS dissipation for the HIT and turbulent channel flows. This concrete evidence urges to go beyond the conventional means of SGS modeling for the scalar flux in order to address the matter of nonlocality with more sophisticated mathematical modeling tools. Thus, a nonlocal construction for the EDM would be a fairly relevant remedy to this problem. 

\section{Boltzmann Transport Framework}\label{sec: BT-framework}

In studying the turbulent transport and mixing, kinetic Boltzmann theory has shown a rich and promising ground based upon principles of statistical mechanics, which by construction is well-suited for the stochastic description of turbulence at microscopic level \citep{harris2004introduction}. In the following, the fundamental sources of nonlocal closure and the SGS modeling for the residual passive scalar flux are studied at the kinetic Boltzmann transport framework. Our objective is to derive a nonlocal eddy-diffusivity SGS model at the continuum level.

\subsection{BGK Model and Double Distribution Function}\label{sec: BGK-closure}

Considering classical kinetic theory of gases, we are concerned with the evolution of a single particle distribution function, $f$, that is governed by the Boltzmann Transport Equation (BTE),
\begin{align}\label{eqn: BTE-fluid}
    \frac{\partial f}{\partial t} + \boldsymbol{u} \cdot \nabla f = C(f).
\end{align}

In \eqref{eqn: BTE-fluid}, the probability distribution $f=f(t,\boldsymbol{x},\boldsymbol{u})$ is defined such that there exists mass of fluid particles that are located inside the infinitesimal volume element $d\boldsymbol{x}$ centered at $\boldsymbol{x}$, velocity element $d\boldsymbol{u}$ centered at $\boldsymbol{u}$, and at time $t$. In the phase space of particle, $\boldsymbol{x}$, $\boldsymbol{u}$, and $t$ are considered as independent variables. The left-hand side of \eqref{eqn: BTE-fluid} represents the streaming of the non-reacting particles that is balanced by the \textit{collision} operator, $C(f)$, on the right-hand side. As a widely common model for the collision operator, Bhatnagar–Gross–Krook (BGK) approximation considers scattering of the fluid particle due to collision with another particle. Therefore, the BGK model characterizes $C(f)=C_{\mathrm{BGK}}(f)$ with a single parameter, that is called the \textit{relaxation time}, $\tau$ \citep{BGK1954}. Therefore, the collision operator is written as
\begin{align}\label{eqn: BGK-coll}
    C_{\mathrm{BGK}}(f) = -\frac{f-f^{eq}}{\tau},
\end{align}
where the local equilibrium distribution function, $f^{eq}=f^{eq}(t,\boldsymbol{x},\boldsymbol{u})$ is given by the Maxwell distribution \citep{sone2012kinetic}, and is parameterized by the locally conserved quantities (density $\rho$, particle speed $\boldsymbol{u}$, and temperature $T$) as
\begin{align}\label{eqn: Maxwell-fluid}
    f^{eq} = \frac{\rho}{(2\pi \, c_T^2)^{d/2}} \exp\left( \frac{-(\boldsymbol{u}-\boldsymbol{V})^2}{2 \, c_T^2} \right).
\end{align}
In \eqref{eqn: Maxwell-fluid}, $c_T=\sqrt{k_B \, T/m}$ is the thermal speed at $T$ in which $k_B$ is the Boltzmann constant, and $m$ represents the molecular air weight, while $d$ denotes the spatial dimensions \citep{huang1987statistical}. 

In order to study the passive scalar transport phenomena in this context, Double Distribution Function (DDF) method has been a successful approach \citep{sharma2020current}. In the DDF, we consider one distribution function to address the conservation of mass and momentum while another distribution function is taken to represent the conservation of energy. In the case of passive scalar transport, the compressive work and heat dissipation are considered to be negligible in the incompressible limit \citep{bartoloni1993lbe, eggels1995numerical, shan1997simulation}. Therefore, the extra BTE that governs the energy distribution function, $g=g(t,\boldsymbol{x},\boldsymbol{u})$, with the BGK collision model is expressed as
\begin{align}\label{eqn: BTE-scalar}
    \frac{\partial g}{\partial t} + \boldsymbol{u} \cdot \nabla g = C_{\mathrm{BGK}}(g) = -\frac{g-g^{eq}}{\tau_g}.
\end{align}
In \eqref{eqn: BTE-scalar}, $\tau_g$ represents the relaxation time, which is the time-scale associated with the collisional relaxation to the local energy equilibrium denoted by the Maxwell energy distribution,
\begin{align}\label{eqn: Maxwell-scalar}
    g^{eq} = \frac{\Phi}{(2\pi \, c_T^2)^{d/2}} \exp\left( \frac{-(\boldsymbol{u}-\boldsymbol{V})^2}{2 \, c_T^2} \right).
\end{align}
Defining $\mathcal{L}=(\boldsymbol{u} - \boldsymbol{V})^2/c_T^2$ and $F(\mathcal{L})=\exp (-\mathcal{L}/2)$, the Maxwell distribution in \eqref{eqn: Maxwell-scalar} (for the most general case where $d=3$) is reformulated as $g^{eq} = \frac{\Phi}{(2\pi)^{3/2} \, c_T^3} \, F(\mathcal{L})$.

Subsequently, continuum averaging yields the macroscopic flow variables for the incompressible flow, $\rho = \rho(t,\boldsymbol{x})$, as follows:
\begin{eqnarray} \label{eqn: continuum-ave1}
    \rho &=& \int_{\mathbb{R}^d} f(t,\boldsymbol{x},\boldsymbol{u}) \, d\boldsymbol{u},
    \\ \label{eqn: continuum-ave2}
    \rho \, \boldsymbol{V}(t,\boldsymbol{x}) &=& \int_{\mathbb{R}^d} \, \boldsymbol{u} \, f(t,\boldsymbol{x},\boldsymbol{u}) \, d\boldsymbol{u}, \qquad i=1,2,3,
    \\ \label{eqn: continuum-ave3}
    \Phi(t,\boldsymbol{x}) &=& \int_{\mathbb{R}^d} g(t,\boldsymbol{x},\boldsymbol{u}) \, d\boldsymbol{u},
\end{eqnarray}
where $\Phi(t,\boldsymbol{x})$ is the total passive scalar concentration field appearing in the AD equation.

Let us define $L$ as the macroscopic characteristic length, $l_s$ as the microscopic characteristic length associated with the smallest length-scale of the passive scalar, and $l_m$ as the mean-free path (the average distance traveled by a particle between successive collisions). Considering $\boldsymbol{x}^{\prime}$ to be the location of particles before scattering while we characterize their current location with $\boldsymbol{x}$, one can assume that $\boldsymbol{x}^{\prime}=\boldsymbol{x}-\delta \boldsymbol{x}$, where $\delta \boldsymbol{x} = (t-t^{\prime}) \, \boldsymbol{u}$. Here we assume that during the time $t-t^{\prime}$, $\boldsymbol{u}$ approximately remains constant \citep{samiee2020fractional}. According to Chen \textit{et al.} \citep{chen2007macroscopic, chen2010macroscopic}, the Boltzmann BGK kinetics with ``constant'' relaxation time, equations \eqref{eqn: BTE-fluid} and \eqref{eqn: BTE-scalar}, admit analytical solutions for $f(t,\boldsymbol{x},\boldsymbol{u})$ and $g(t,\boldsymbol{x},\boldsymbol{u})$ based upon their local equilibrium distribution that is valid in a general flow where the distance from the wall is large compared to $l_m$. Focusing on equation \eqref{eqn: BTE-scalar} and defining $s = (t-t^\prime)/\tau_g$, the exact solution to $g(t,\boldsymbol{x},\boldsymbol{u})$ would be
\begin{align}\label{eqn: analytic-sol-g}
    g(t,\boldsymbol{x},\boldsymbol{u}) = \int_{0}^{\infty} e^{-s} \, g^{eq}(t-s\tau_g, \, \boldsymbol{x}- \boldsymbol{u} \, s\tau_g, \, \boldsymbol{u}) \, ds
    =\int_{0}^{\infty} e^{-s} \, g^{eq}_{s,s}(\mathcal{L}) \, ds,
\end{align}
where $g^{eq}_{s,s}(\mathcal{L})=g^{eq}(t-s\tau_g, \, \boldsymbol{x}-\boldsymbol{u} \, s\tau_g, \, \boldsymbol{u})$. 

\subsection{Filtered BTE, Closure Problem, and Kinetic-Boltzmann Modeling}\label{sec: FBTE}

Statistical description of LES is well-represented through incorporating a filtering procedure into the kinetic Boltzmann transport. For the purpose of passive scalar transport, applying a spatially and temporally invariant filtering kernel, $\boldsymbol{\mathcal{G}} = \boldsymbol{\mathcal{G}}(\boldsymbol{r})$, onto the distribution  function $g(t,\boldsymbol{x},\boldsymbol{u})$ linearly decomposes that into the filtered, $\widetilde{g}=\boldsymbol{\mathcal{G}} \ast g$, and the residual, $g^{\prime}=g-\widetilde{g}$, components. Therefore, filtering the equation \eqref{eqn: BTE-scalar} results in the following filtered BTE (FBTE) for the passive scalar:
\begin{equation}\label{eqn: FBTE}
\frac{\partial \widetilde{g}}{\partial t} + \boldsymbol{u}\cdot \nabla \, \widetilde{g} = -\frac{\widetilde{g}-\widetilde{g^{eq}(\mathcal{L})}}{\tau_g}.
\end{equation}
As it was elaborated by Girimaji \citep{girimaji2007boltzmann}, the nonlinear nature of the collision operator, $C_{\mathrm{BGK}}(g)$, prohibits the filtering kernel to commute with $C_{\mathrm{BGK}}(g)$; thus, it initiates a source of closure at the kinetic level in FBTE \eqref{eqn: FBTE}. Defining $\widetilde{\mathcal{L}}:=(\boldsymbol{u}-\widetilde{\boldsymbol{V}})^2/c_T^2$, this closure problem is manifested in the following inequality,
\begin{align}\label{eqn: kinetic-closure}
    \widetilde{g^{eq}(\mathcal{L})} = \frac{\reallywidetilde{\Phi \, \exp(-\mathcal{L}/2)}}{(2\pi)^{3/2} \, c_T^3} \neq \frac{\widetilde{\Phi} \, \exp(-\widetilde{\mathcal{L}}/2)}{ (2\pi)^{3/2} \, c_T^3} = g^{eq}(\widetilde{\mathcal{L}}).
\end{align}
The identified closure requires proper means of modeling so that one can numerically solve the FBTE \eqref{eqn: FBTE}. A common practice is to approximate this closure problem with a modified relaxation time approach that is described in detail in \citep{sagaut2010toward}. Despite the success of this approach in some applications, it is not physically consistent with the filtered turbulent transport dynamics \citep{girimaji2007boltzmann}. Nevertheless, here we manage to adjust this inconsistency by looking at the nonlocal effects arising from filtering the Maxwell distribution function, $g^{eq}(\mathcal{L})$, and model them with proper mathematical tools. Considering the spatial filtering kernel $\boldsymbol{\mathcal{G}}(\boldsymbol{r})$ with the filter-width $\Delta$, and applying it on the Maxwell equilibrium distribution as
\begin{equation}\label{eqn: F-Maxwell}
    \widetilde{g^{eq}(\mathcal{L})} =  \boldsymbol{\mathcal{G}} \ast g^{eq}\big (\mathcal{L}(t,\boldsymbol{u},\boldsymbol{x})\big ) =  \int_{R_f}^{} \boldsymbol{\mathcal{G}}(\boldsymbol{r}) \,  g^{eq}\big (\mathcal{L}(t,\boldsymbol{u},\boldsymbol{x}-\boldsymbol{r})\big ) \, d\boldsymbol{r},
\end{equation}
where $R_f=[-\Delta/2 \, , \Delta/2]^3$. 

\begin{remb}
The integral form of the convolution \eqref{eqn: F-Maxwell} implies that $\widetilde{g^{eq}(\mathcal{L})}$ consists of a summation of the exponential functions. Thus, filtering encodes a multi-exponential behavior into the filtered equilibrium distribution that is gets intensified as the filter-width enlarges. Moreover, this  multi-exponential structure of the filtered Maxwell distribution induces a heavy-tailed form for the filtered distribution that essentially entails the non-Gaussian behavior and justifies the spatial nonlocality \citep{samiee2020fractional}. This statistical rationale strongly indicates that modeling this closure problem with a Gaussian-type distribution is fundamentally insufficient. On the other hand, it is well-known that the statistical behavior of a multi-exponential distribution could be sufficiently approximated with a power-law distribution \citep{chu2010power, samiee2020fractional}. 
\end{remb}

Subsequently, by rewriting the right-hand side of the passive scalar FBTE \eqref{eqn: FBTE} into the following form
\begin{align}\label{eqn: FBTE-RHS}
     -\frac{1}{\tau_g} \left(\widetilde{g} - \widetilde{g^{eq}(\mathcal{L})} \right) = \underbrace{-\frac{1}{\tau_g} \left(\widetilde{g} - g^{eq}(\widetilde{\mathcal{L}}) \right)}_{\text{closed}} + \underbrace{\frac{1}{\tau_g} \left(\widetilde{g^{eq}(\mathcal{L})} - g^{eq}(\widetilde{\mathcal{L}}) \right)}_{\text{unclosed}},
\end{align}
the unclosed part is structurally multi-exponentially distributed and maybe approximated by a power-law distribution model as we propose 
\begin{align}\label{eqn: Levy-model}
     \widetilde{g^{eq}(\mathcal{L})} - g^{eq}(\widetilde{\mathcal{L}}) \approx g^\alpha(\widetilde{\mathcal{L}}) = \frac{\widetilde{\Phi}}{c_T^3} \, F^{\alpha}(\widetilde{\mathcal{L}}),
\end{align}
where $F^{\alpha}(\widetilde{\mathcal{L}})$ denotes an $\alpha$-stable L\'evy distribution that is mathematically designed based on heavy-tailed stochastic processes and replicate the power-law behavior \citep{applebaum2009levy, meerschaert2011stochastic}. 

Regarding the decomposition given in \eqref{eqn: FBTE-RHS}, and by applying the filtering kernel on the analytical solution to $g(t,\boldsymbol{x}, \boldsymbol{u})$ that is given in \eqref{eqn: analytic-sol-g}, we obtain
\begin{align}\label{eqn: filtered-analytic-g1}
    \widetilde{g} (t,\boldsymbol{x}, \boldsymbol{u}) = \int_{0}^{\infty} e^{-s} \,  \widetilde{g^{eq}_{s,s}(\mathcal{L})}\, ds
    = \int_{0}^{\infty} e^{-s} \,  g^{eq}_{s,s}(\widetilde{\mathcal{L}}) \, ds + \int_{0}^{\infty} e^{-s} \,  \left(\widetilde{g^{eq}_{s,s}(\mathcal{L})} - g^{eq}_{s,s}(\widetilde{\mathcal{L}})\right) \, ds,
\end{align}
where $\widetilde{g^{eq}_{s,s}(\mathcal{L})}=\reallywidetilde{g^{eq}\big (\mathcal{L}(t-s\tau_g,\boldsymbol{x}- \boldsymbol{u} \, s\tau_g,\boldsymbol{u}) \big )}$, and the second integral represents the closure source. Therefore, employing the power-law distribution model in \eqref{eqn: Levy-model} returns the following analytic form for $\widetilde{g}(t,\boldsymbol{x}, \boldsymbol{u})$
\begin{align}\label{eqn: filtered-analytic-g2}
    \widetilde{g} (t,\boldsymbol{x}, \boldsymbol{u}) = \int_{0}^{\infty} e^{-s} \,  g^{eq}_{s,s}(\widetilde{\mathcal{L}}) \, ds + \int_{0}^{\infty} e^{-s} \,  g^{\alpha}_{s,s}(\widetilde{\mathcal{L}}) \, ds,
\end{align}
wherein, $g^{\alpha}_{s,s}(\widetilde{\mathcal{L}}):=g^{\alpha}\big (\widetilde{\mathcal{L}}(t-s\tau_g,\boldsymbol{x}- \boldsymbol{u} \, s\tau_g,\boldsymbol{u}) \big )$.
\subsection{Fractional-Order Model for the SGS Scalar Flux}\label{sec: Derivation}

Similar to the continuum averaging shown in \eqref{eqn: continuum-ave1} to \eqref{eqn: continuum-ave3}, the macroscopic continuum variables associated with \eqref{eqn: Filtered-AD}, are obtained in terms of the filtered distribution functions, $\widetilde{f}$ and $\widetilde{g}$, as
\begin{eqnarray}\label{eqn: continuum-ave-fPhi}
    \widetilde{\Phi} &=& \int_{\mathbb{R}^d} \widetilde{g}(t,\boldsymbol{x},\boldsymbol{u}) \,  d\boldsymbol{u},
    \\ \label{eqn: continuum-ave-fVPhi}
    \widetilde{V}_i &=& \frac{1}{\rho}\int_{\mathbb{R}^d} u_i \, \widetilde{f}(t,\boldsymbol{x},\boldsymbol{u}) \, d\boldsymbol{u}, \quad i=1,2,3.
\end{eqnarray}
Multiplying both sides of the passive scalar FBTE by a collisional invariant $\mathcal{X}=\mathcal{X}(\boldsymbol{u})$ and then integrating over the kinetic momentum would return 
\begin{equation}\label{eqn: FBTE-ave-general}
    \int_{\mathbb{R}^d} \mathcal{X} \left( \frac{\partial \widetilde{g}}{\partial t} + \boldsymbol{u}\cdot \nabla \, \widetilde{g}\right) d\boldsymbol{u} = \int_{\mathbb{R}^d} \mathcal{X} \left( -\frac{\widetilde{g}-\widetilde{g(\mathcal{L})}}{\tau_g} \right) d\boldsymbol{u}.
\end{equation}
Here, choosing $\mathcal{X} = 1$ would result in recovering the filtered AD equation \eqref{eqn: Filtered-AD}. According to the microscopic reversibility of the particles that assumes the collisions occur \textit{elastically}, the right-hand side of \eqref{eqn: FBTE-ave-general} equals zero \citep{saint2009hydrodynamic}. Therefore, \eqref{eqn: FBTE-ave-general} reads as
\begin{align}\label{GE-14}
    \frac{\partial \widetilde{\Phi}}{\partial t} + \nabla\cdot \int_{\mathbb{R}^d} \boldsymbol{u} \, \widetilde{g} \, d\boldsymbol{u} = 0. 
\end{align}
Since we are working with spatial filtering kernels, $\boldsymbol{\mathcal{G}}=\boldsymbol{\mathcal{G}}(\boldsymbol{r})$,
\begin{align}\label{GE-16}
    \int_{\mathbb{R}^d} \boldsymbol{u} \, \widetilde{g} \, d\boldsymbol{u} = \int_{\mathbb{R}^d} (\boldsymbol{u}-\widetilde{\boldsymbol{V}}) \, \widetilde{g} \, d\boldsymbol{u}+ \int_{\mathbb{R}^d} \widetilde{\boldsymbol{V}} \, \widetilde{g}\, d\boldsymbol{u}.
\end{align}
By plugging \eqref{GE-16} into \eqref{GE-14}, we obtain that
\begin{equation}\label{GE-17}
    \frac{\partial \widetilde{\Phi}}{\partial t} + \nabla \cdot \left(\widetilde{\Phi} \, \widetilde{\boldsymbol{V}}\right) = -\nabla \cdot \boldsymbol{q},
\end{equation}
where 
\begin{equation}\label{GE_17_2}
    q_i=\int_{\mathbb{R}^d} \left(u_i-\widetilde{V}_i\right) \, \widetilde{g} \, d\boldsymbol{u}.
\end{equation} 
Using \eqref{eqn: filtered-analytic-g2}, we formulate $q_i$ as 
\begin{align}\label{GE-19-1}
    q_i = \int_{\mathbb{R}^d}\int_{0}^{\infty} e^{-s} (u_i-\widetilde{V}_i) \, g^{eq}_{s,s}(\widetilde{\mathcal{L}}) \, ds \, d\boldsymbol{u} + \int_{\mathbb{R}^d}\int_{0}^{\infty} e^{-s} (u_i-\widetilde{V}_i) \, g^{\alpha}_{s,s}(\widetilde{\mathcal{L}}) \, ds \, d\boldsymbol{u}.
\end{align}
It is straightforward to show that the temporal shift can be removed from \eqref{GE-19-1}. Moreover, since $(u_i-\widetilde{V}_i) \, g^{eq}(\widetilde{\mathcal{L}})$ and $(u_i-\widetilde{V}_i) \, g^{eq}(\widetilde{\mathcal{L}})$ both represent odd functions of $u_i$, thus,  
\begin{align}
    \int_{\mathbb{R}^d}(u_i-\widetilde{V}_i) \, g^{eq}(\widetilde{\mathcal{L}}) \, d\boldsymbol{u} = \int_{\mathbb{R}^d}(u_i-\widetilde{V}_i) \, g^{\alpha}(\widetilde{\mathcal{L}}) \, d\boldsymbol{u} = 0.
\end{align}
As a result, $q_i$ in \eqref{GE-19-1} can be rewritten as
\begin{align}\label{GE-20}
    q_i &=& \int_{\mathbb{R}^d}\int_{0}^{\infty} e^{-s} (u_i-\widetilde{V}_i) \left(g^{eq}_{s,s}(\widetilde{\mathcal{L}}) - g^{eq}(\widetilde{\mathcal{L}})\right) ds \, d\boldsymbol{u} \, + \int_{\mathbb{R}^d}\int_{0}^{\infty} e^{-s} (u_i-\widetilde{V}_i) \left(g^{\alpha}_{s,s}(\widetilde{\mathcal{L}}) - g^{\alpha}(\widetilde{\mathcal{L}})\right) ds \, d\boldsymbol{u}.
\end{align}
In an LES setting, the first integral on the right-hand side of \eqref{GE-20} represents the \textit{filtered} scalar flux, $\widetilde{\boldsymbol{q}}$, while the second integral aims to model the \textit{residual} scalar flux, $\boldsymbol{q}^R$, associated with unresolved small scales of turbulent transport. In other words, by assigning the Gaussian distribution $g^{eq}(\widetilde{\mathcal{L}})$ to $\widetilde{q_i}$ and the isotropic $\alpha$-stable L\'evy distribution, $g^{\alpha}(\widetilde{\mathcal{L}})$, to $q_i^R$, the total passive scalar flux, $\boldsymbol{q}=\widetilde{\boldsymbol{q}}+\boldsymbol{q}^R$, in \eqref{GE-20} may be decomposed as
\begin{eqnarray}\label{GE-25}
    \widetilde{q_i} &=& \int_{0}^{\infty}  \int_{\mathbb{R}^d} (u_i-\widetilde{V}_i) \left(g^{eq}_{s,s}(\widetilde{\mathcal{L}})-g^{eq}(\widetilde{\mathcal{L}})\right) e^{-s} d\boldsymbol{u} \, ds,
    \\
    \label{GE-26}
    q_i^R &=& \int_{0}^{\infty}  \int_{\mathbb{R}^d} (u_i-\widetilde{V}_i) \left(g^{\alpha}_{s}(\widetilde{\mathcal{L}})-g^{\alpha}(\widetilde{\mathcal{L}})\right) e^{-s} d\boldsymbol{u} \, ds.
\end{eqnarray}

In \ref{sec: Appendix1}, the details of derivation of $\widetilde{\boldsymbol{q}}$ and $\boldsymbol{q}^R$ in terms of macroscopic transport variables including $\widetilde{\Phi}$ and $\widetilde{\boldsymbol{V}}$ are presented. As the result, the filtered passive scalar flux is obtained as
\begin{align}\label{eqn: flt-flux}
    \widetilde{\boldsymbol{q}} = -\mathcal{D} \, \nabla \widetilde{\Phi},
\end{align}
and the divergence of residual scalar flux is derived as the fractional Laplacian of the filtered total scalar concentration,
\begin{align}\label{eqn: res-flux}
    \nabla \cdot \boldsymbol{q}^R = -\mathcal{D}_\alpha \, (-\Delta)^{\alpha} \, \widetilde{\Phi}, \quad \alpha \in (0,1],
\end{align}
where $\mathcal{D}_\alpha := \frac{C_\alpha (c_T \, \tau_g)^{2\alpha}}{\tau_g} \, (2\alpha+2) \, \Gamma(2\alpha)$ is a model coefficient with the unit [$L^{2\alpha}/T$]. The filtered AD equation for the total passive scalar concentration, developed from the filtered kinetic BTE with an $\alpha$-stable L\'evy distribution model, yields a fractional-order SGS scalar flux model at the continuum level. The aforementioned filtered AD equation reads as
\begin{align} \label{eqn: Flt-AD-total}
    \frac{\partial \widetilde{\Phi}}{\partial t}+\frac{\partial}{\partial x_i}\left( \widetilde{\Phi} \, \widetilde{V}_i\right) = \mathcal{D} \, \Delta \widetilde{\Phi} +\mathcal{D}_{\alpha} (-\Delta)^{\alpha} \, \widetilde{\Phi}.
\end{align}
Through a proper choice for the fractional Laplacian order $\alpha$, the developed model optimally works in an LES setting. Applying the Reynolds decomposition and considering the passive scalar with imposed uniform mean gradient, equation \eqref{eqn: Flt-AD-total} fully recovers the filtered transport equation \eqref{eqn: Filtered-AD} for the transport of the filtered scalar fluctuations, $\widetilde{\phi}$.

In order to explicitly derive the modeled residual scalar flux in terms of the filtered transport fields, from the Fourier definition of fractional Laplacian and the Riesz transform in given in \ref{sec: Fractional-Calc}, one can verify that 
\begin{eqnarray}
\mathcal{F} \Big {\{} (-\Delta)^{\alpha} \, \widetilde{\phi} \Big {\}} =  \mathfrak{i} \, \xi_j \Big ( -\mathfrak{i} \, \xi_j / \vert \boldsymbol{\xi} \vert \Big) \, (\vert \boldsymbol{\xi} \vert^2 )^{\alpha-\frac{1}{2}} \, \mathcal{F} \Big {\{} \widetilde{\phi} \Big {\}},
\end{eqnarray}
which leads to
\begin{equation}\label{Flx-1}
    (-\Delta)^{\alpha} \widetilde{\phi} = \nabla_j \left(\mathcal{R}_j (-\Delta)^{\alpha-\frac{1}{2}} \, \widetilde{\phi}\right).
\end{equation}
Therefore, using \eqref{eqn: res-flux} we may write
\begin{equation}\label{Flx-1-2}
    \nabla \cdot \boldsymbol{q}^{R} =  \nabla \cdot \left(-\mathcal{D}_\alpha \, \mathcal{R} (-\Delta)^{\alpha-\frac{1}{2}} \, \widetilde{\phi}\right).
\end{equation} 
Finally, from \eqref{Flx-1-2} one can find the \textit{explicit} form of the modeled SGS flux as
\begin{equation}\label{Flx-2}
    q^{R}_{i} = -\mathcal{D}_\alpha \, \mathcal{R}_i (-\Delta)^{\alpha-\frac{1}{2}} \, \widetilde{\phi} + c,
\end{equation}
where $c$ is a real-valued constant. 

\section{Data-driven Nonlocal SGS Modeling}\label{sec: Calibration}

Deriving the structure of the residual scalar flux as a nonlocal SGS model, there are two levels of model calibration in order to employ this SGS model in an LES. In fact, this model calibration problem could be viewed as a two-stage procedure where its first part is dealing with estimation of the fractional order, $\alpha$, and the other stage infers the proportionality coefficient of the model, $\mathcal{D}_\alpha$. Subsequently, we propose a two-stage \textit{a priori} parameter identification strategy based upon spatio-temporal data for the true $\boldsymbol{q}^R$, obtained from filtering well-resolved DNS of scalar turbulence as described in section \ref{sec: nonlocality}.

\subsection{Capturing Nonlocality with Fractional Modeling of the SGS Scalar Flux}\label{sec: TPCs}

This is the first stage of this data-driven model identification, which targets finding an optimal fractional order, $\alpha_{opt}$. Our ground-truth data comes from exact evaluation of the two-point correlation function, $\mathcal{C}(q^R_\parallel \, , \widetilde{G}_\parallel)$ as described in section \ref{sec: nonlocality}. In fact, we aim to capture the spatial nonlocality we showed in the statistics of SGS production of filtered scalar variance (see Figure \ref{fig: Nonlocality}b). Since we employ the fluctuating part of $q^R_\parallel$ in computing the two-point correlation function, and from the definition $\mathcal{C}(q^R_\parallel \, , \widetilde{G}_\parallel)$ is normalized by $\left\langle q^R_\parallel(\boldsymbol{x}) \, \widetilde{G}_\parallel(\boldsymbol{x}) \right\rangle$, finding $\alpha_{opt}$ is essentially independent of the other model parameters appeared in \eqref{Flx-2}. Using the exact values of $q^R_\parallel$ from filtered DNS, \eqref{eqn: TPC1} returns the ground-truth two-point correlation function, $\mathcal{C}^\mathrm{True}$ Using the database described in section \ref{sec: nonlocality}, while using the fractional model for SGS scalar return flux $\mathcal{C}^\mathrm{Model}$ as functions of spatial shift, $r$. In our study, for a fixed filter width, the fractional order that minimizes the mismatch function $\Vert \mathcal{C}^\mathrm{True} - \mathcal{C}^\mathrm{Model}\Vert$, simply determines $\alpha_{opt}$ capturing the entire range of spatial nonlocality.

\begin{figure}[t!]
    \begin{minipage}[b]{.49\linewidth}
        \centering
        \includegraphics[width=1\textwidth]{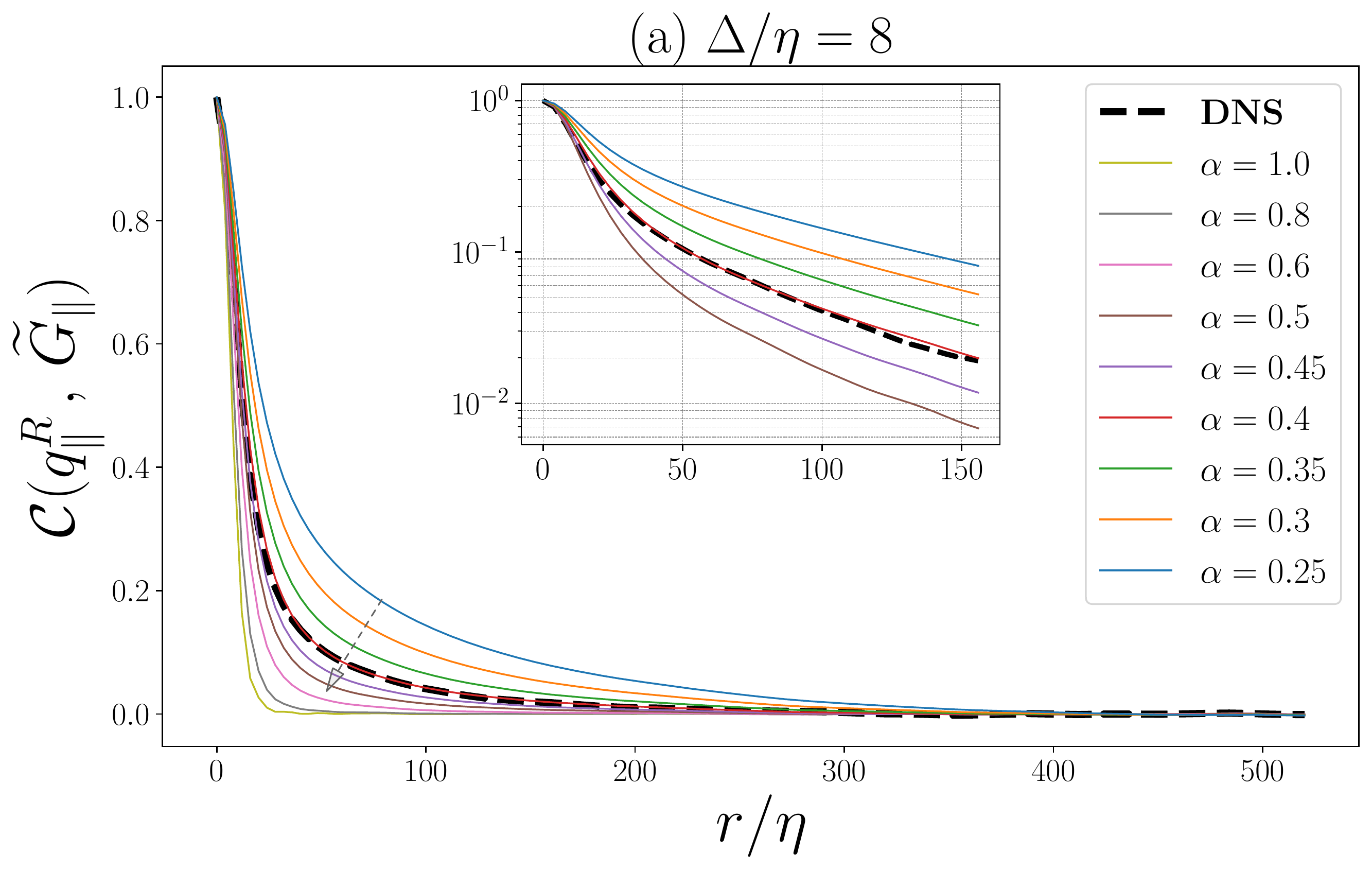}
    \end{minipage}
    \begin{minipage}[b]{.02\linewidth}
    		~
    \end{minipage}
    \begin{minipage}[b]{.49\linewidth}
        \centering
        \includegraphics[width=1\textwidth]{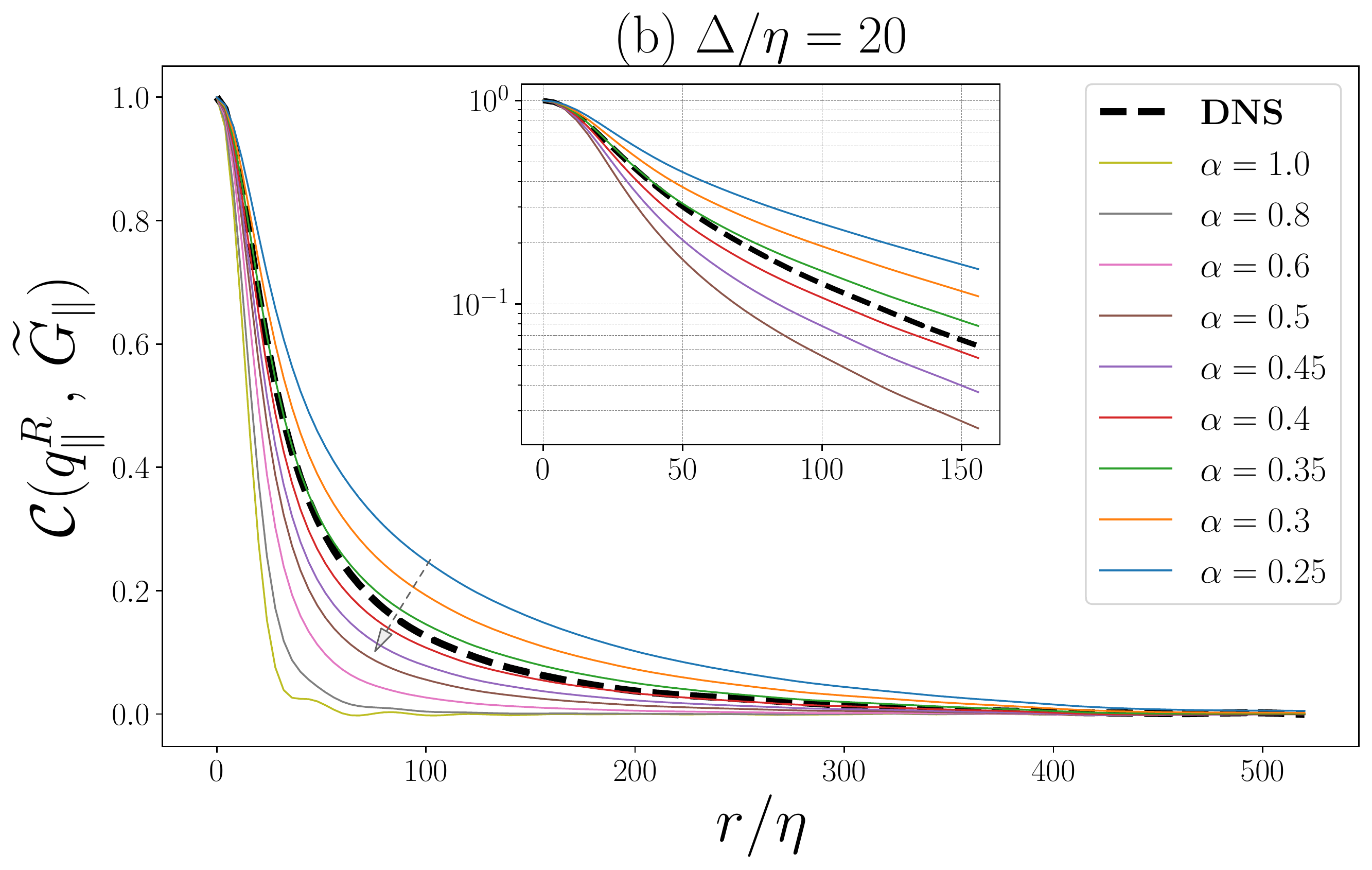}
    \end{minipage}

	\begin{minipage}[b]{1\linewidth}
    		~
    \end{minipage}
    \begin{minipage}[b]{.49\linewidth}
        \centering
        \includegraphics[width=1\textwidth]{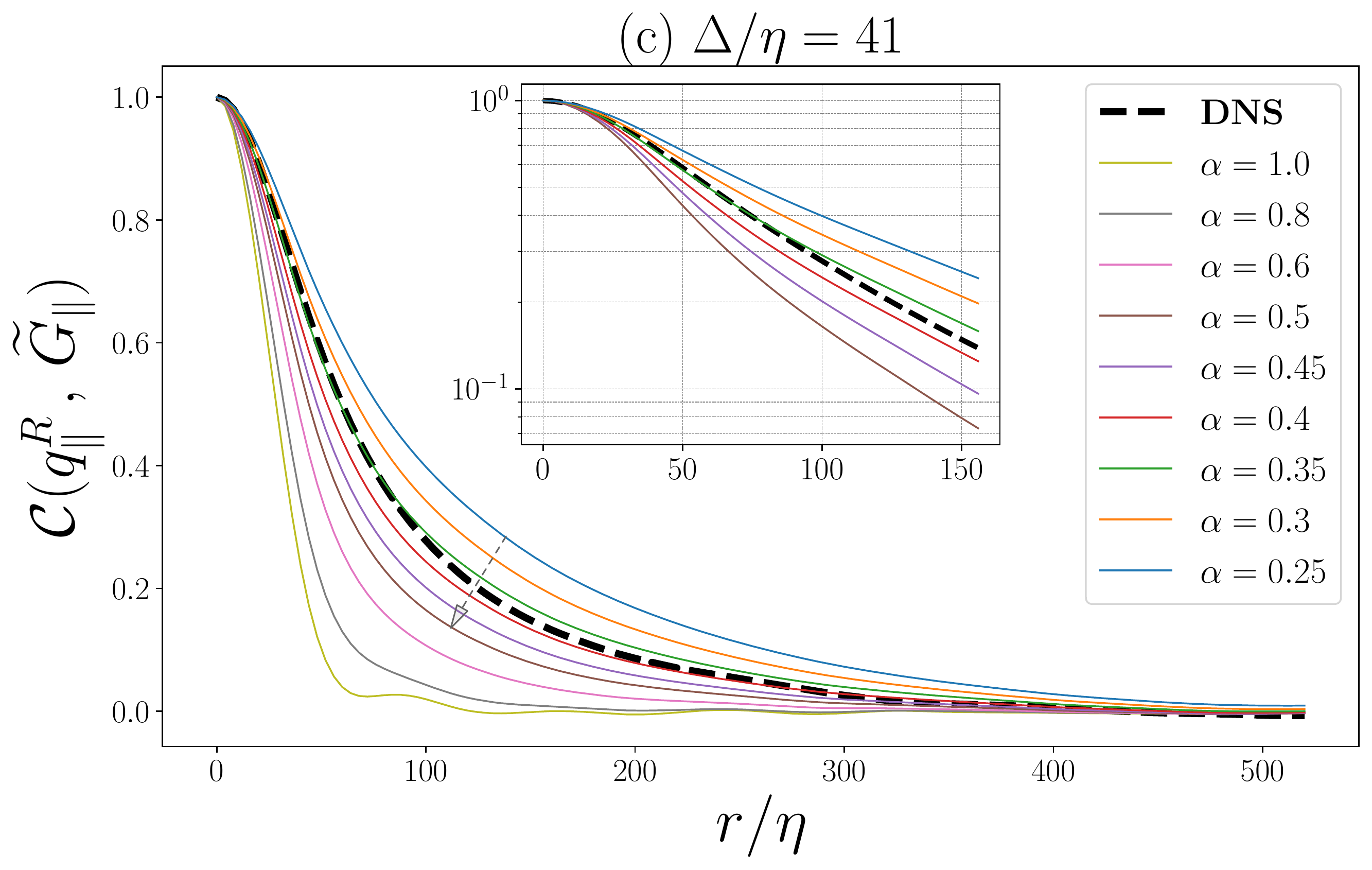}
    \end{minipage}
    \begin{minipage}[b]{.02\linewidth}
    		~
    \end{minipage}
    \begin{minipage}[b]{.49\linewidth}
        \centering
        \includegraphics[width=1\textwidth]{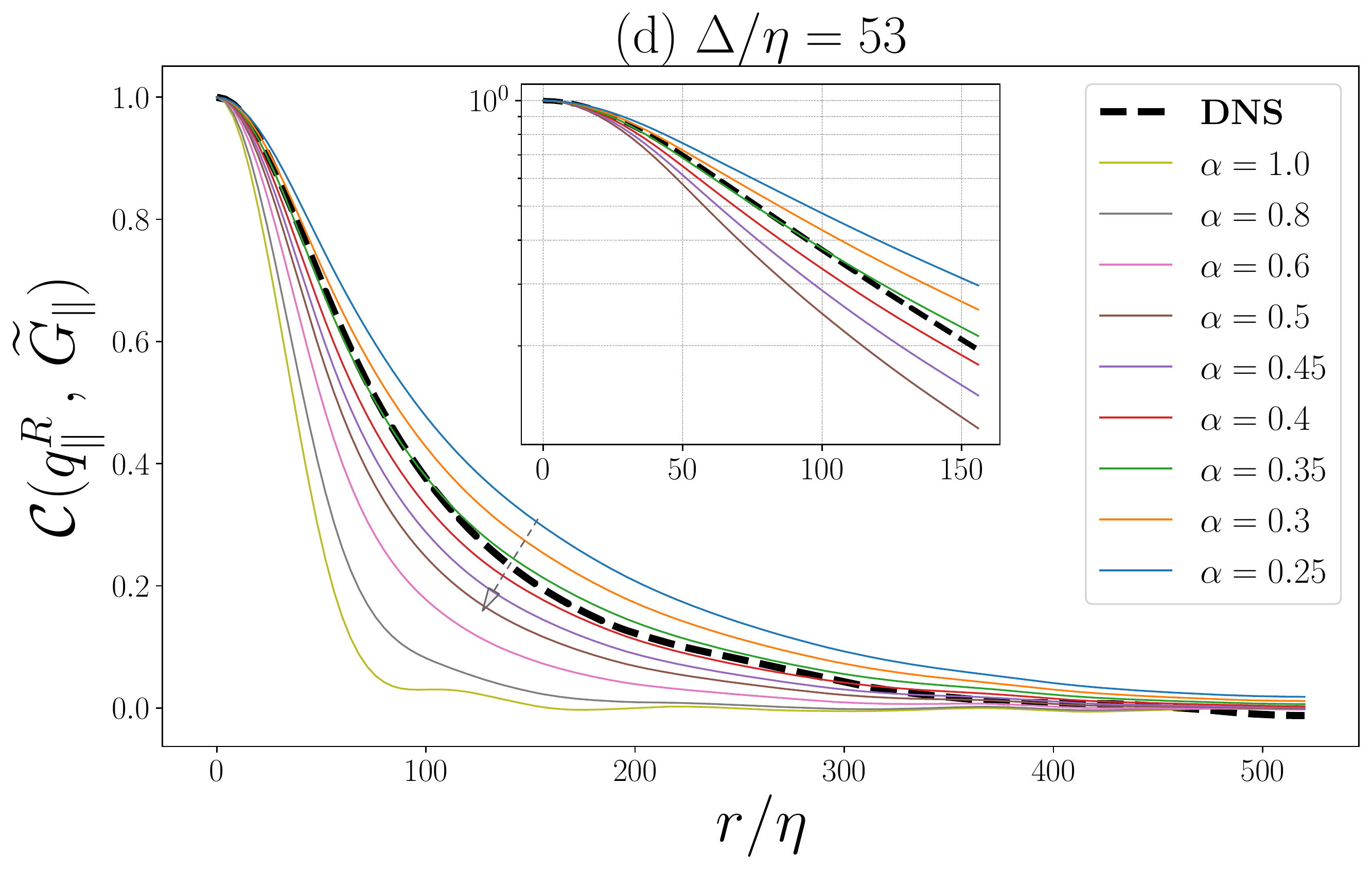}
    \end{minipage}     
    \caption{\footnotesize  Variations of the the two-point correlation function given in \eqref{eqn: TPC1} obtained for the modeled SGS flux $\mathcal{C}^{\mathrm{Model}}$ $\alpha$ is changed from 0 to 1, in addition to the exact evaluation of $\mathcal{C}(q^R_\parallel \, , \, \widetilde{G}_\parallel)$ via exact values of the SGS scalar flux from DNS, illustrated for four filter widths (a) $\Delta/\eta =8$, (b) $\Delta/\eta =20$, (c) $\Delta/\eta =41$, and (d) $\Delta/\eta =53$. The arrows indicate the increase of $\alpha$. Insets depict the two-point correlation function values on smaller regions of the spatial shift, $r/\eta < 150$, in logarithmic scale. These plots show that the true values of the two-point correlation function over the entire range of spatial shift is well-approximated with finding the $\alpha_{opt}$ in the fractional-order SGS model.}\label{fig: optimal_alpha}
\end{figure}

By changing $0 < \alpha \leq 1$, we evaluate $\mathcal{C}^\mathrm{Model}$ for four different filter widths, $\Delta/\eta = 8, \, 20, \, 41, \, 53$. Figure \ref{fig: optimal_alpha}, shows $\mathcal{C}^\mathrm{True}$ in addition to the variations of $\mathcal{C}^\mathrm{Model}$ with $r/\eta$ as we change $\alpha$. We observe that as $\alpha$ decreases, the nonlocal correlations in $\mathcal{C}^\mathrm{True}$ are better approximated over $r$ with the fractional SGS model. According to the minimization of the mismatch function we introduced, $\alpha_{opt}$ for the four values of filter width is reported in Table \ref{tab: one-point}. Moreover, given $\alpha_{opt}$ for each filter width, single-point correlation coefficient between the true and modeled values of the SGS scalar flux, $\varrho \left(q_{\parallel}^{\mathrm{True}}, \, q_{\parallel}^{\mathrm{Model}}\right)$, is computed and acceptably good correlation values (in an \textit{a priori} sense) are reported in Table \ref{tab: one-point}. We need to emphasize that the passive scalar transport occurs in a statistically homogeneous medium with a direction of large-scale anisotropy. This source of anisotropy significantly impacts the intensity of nonlocal effects in the SGS dynamics so that the identified fractional-order in the SGS model is found to be less than 0.5. A similar observation in the study by Di Leoni \textit{et al.} showed that presence of anisotropy effects in the turbulent channel flow (due to the non-zero mean velocity gradient along the stream-wise direction) increases the nonlocality in the SGS dynamics in a way that it requires $\alpha < 0.5$ to properly capture that with the fractional gradient SGS model \citep{di2020two}.

\begin{table}[t!]
	\caption{\footnotesize Optimal fractional orders, and their corresponding single-point correlation coefficients between true and modeled SGS scalar fluxes.}\label{tab: one-point}
	\centering
	\begin{tabular}{ccccc}
			\toprule \toprule
			$\Delta/\eta$     &  $\quad$ & $\alpha_{opt}$    &  $\quad$ & $\varrho \left(q_{\parallel}^{\mathrm{True}}, \, q_{\parallel}^{\mathrm{Model}}\right)$ \\
		\midrule
		8   &  $\quad$  & 0.40  &  $\quad$  & 0.35 \\
		20 &  $\quad$ & 0.35  &  $\quad$  & 0.40\\
		41  &  $\quad$  & 0.36  &  $\quad$  & 0.44\\
		53 &  $\quad$ & 0.37  &  $\quad$  & 0.45\\
		\bottomrule \bottomrule 
	\end{tabular}
\end{table}

\subsection{Sparse Regression on the Fractional-Order Model}\label{sec: regression}

After obtaining the $\alpha_{opt}$ for a choice of filter width, we can compute the explicit term $\boldsymbol{X}=\mathcal{R}(-\Delta)^{\alpha_{opt}-\frac{1}{2}}\, \widetilde{\phi}$ noting the linear mapping $\boldsymbol{q}^R = - \mathcal{D}_\alpha \, \boldsymbol{X}+c$, in \eqref{Flx-2}. Having access to the true values of SGS scalar flux on an extensive spatio-temporal database (described in section \ref{sec: nonlocality}) turns the second stage of our model calibration into a \textit{sparse linear regression} procedure. Therefore, this procedure leads to learning and inferring of $\mathcal{D}_\alpha$ that is appeared in the filtered AD equation \eqref{eqn: Flt-AD-total}.

\begin{figure}[t!]
    \begin{minipage}[b]{.49\linewidth}
        \centering
        \includegraphics[width=1\textwidth]{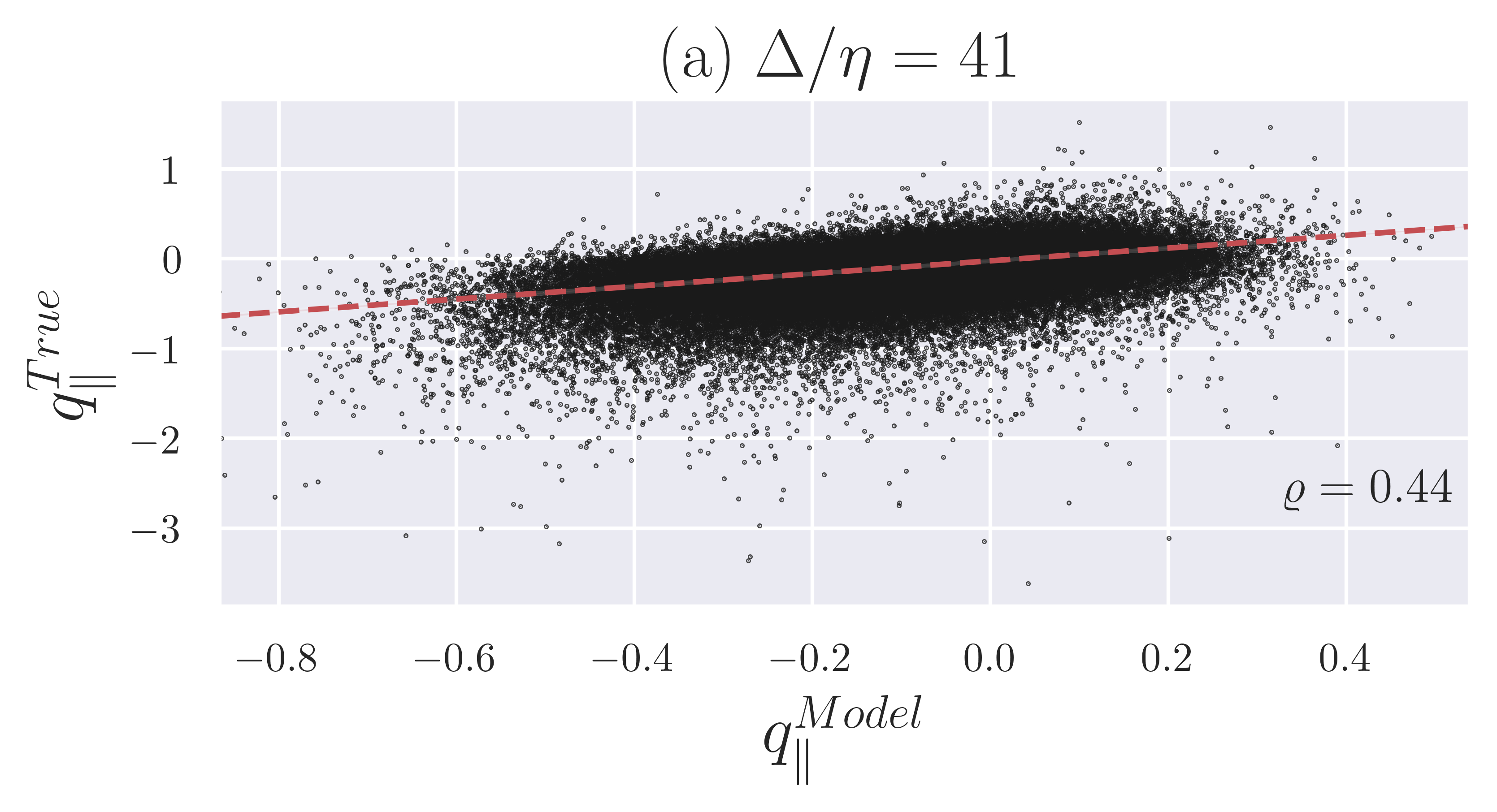}
    \end{minipage}
    \begin{minipage}[b]{.02\linewidth}
    		~
     \end{minipage}
    \begin{minipage}[b]{.49\linewidth}
        \centering
        \includegraphics[width=1\textwidth]{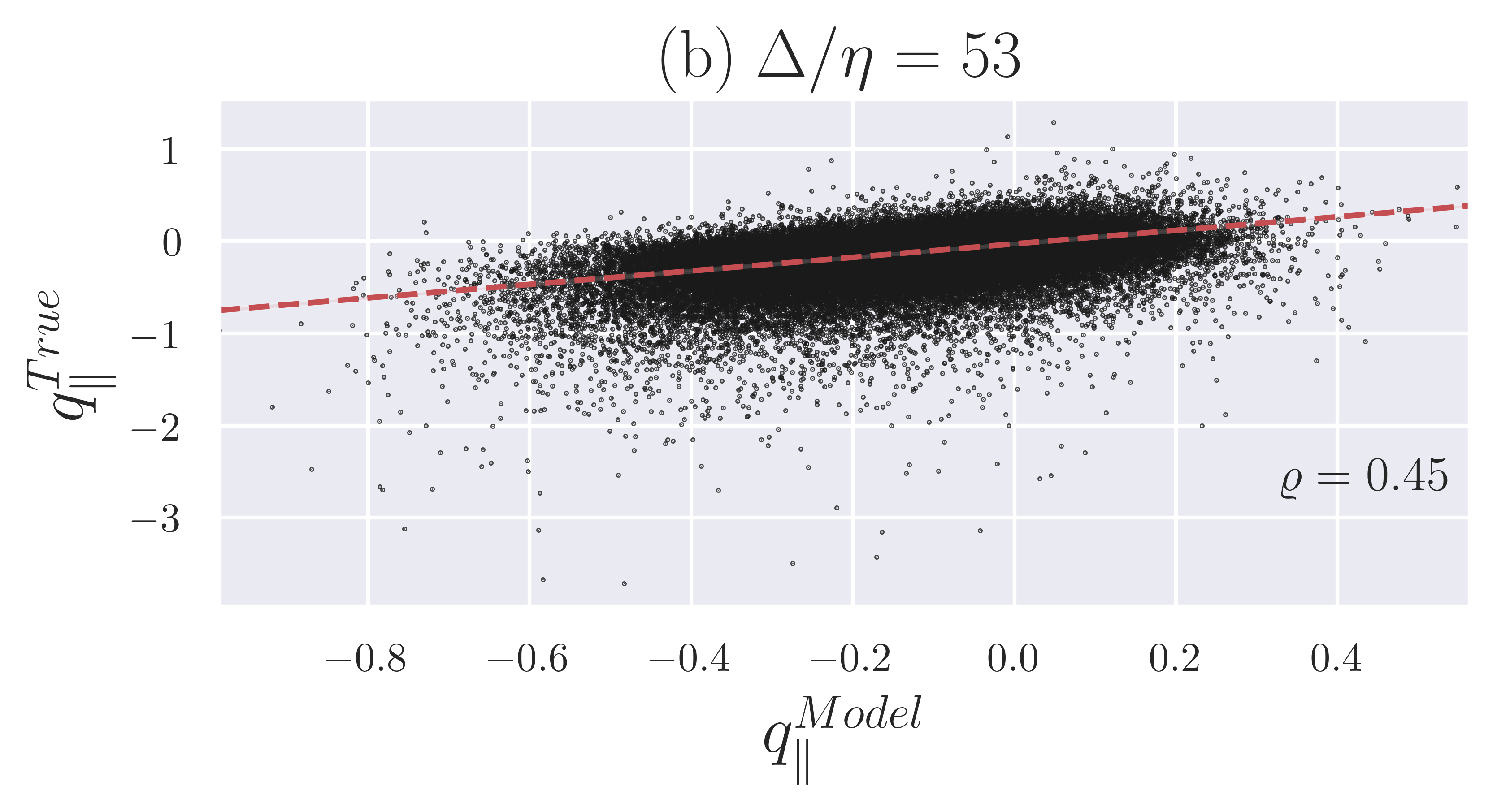}
    \end{minipage}   
    \caption{\footnotesize Regression plots between $q^{\mathrm{True}}_{\parallel}$ and $q^{\mathrm{Model}}_{\parallel}$ for the filter widths, (a) $\Delta/\eta=41$, and (b) $\Delta/\eta=53$. The corresponding optimal fractional-orders are reported in Table \ref{tab: one-point}.}\label{fig: regression}
\end{figure}

Similar to Beetham and Capecelatro's work for sparse regression \citep{beetham2020formulating}, we employ a regularized linear regression method namely as \textit{elastic net} that combines the $L_1$ and $L_2$ penalties as its regularizer \citep{zou2005regularization}. Using the implementation of the elastic net method in \texttt{scikit-learn} \citep{scikit-learn} and assigning equal weights to the $L_1$ and $L_2$ regularizes, we perform the regression and the its quality is examined through scatter plots. As a common practice, and in order to choose proper training data size, we perform cross-validation tests over our spatio-temporal dataset \citep{kutz2017deep}. As a result, Figure \ref{fig: regression} shows the resulting scatter plots after the regression for two cases with $\Delta/\eta=41, \, 53$.

\begin{figure}[t!]
    \centering
    \includegraphics[width=.5\textwidth]{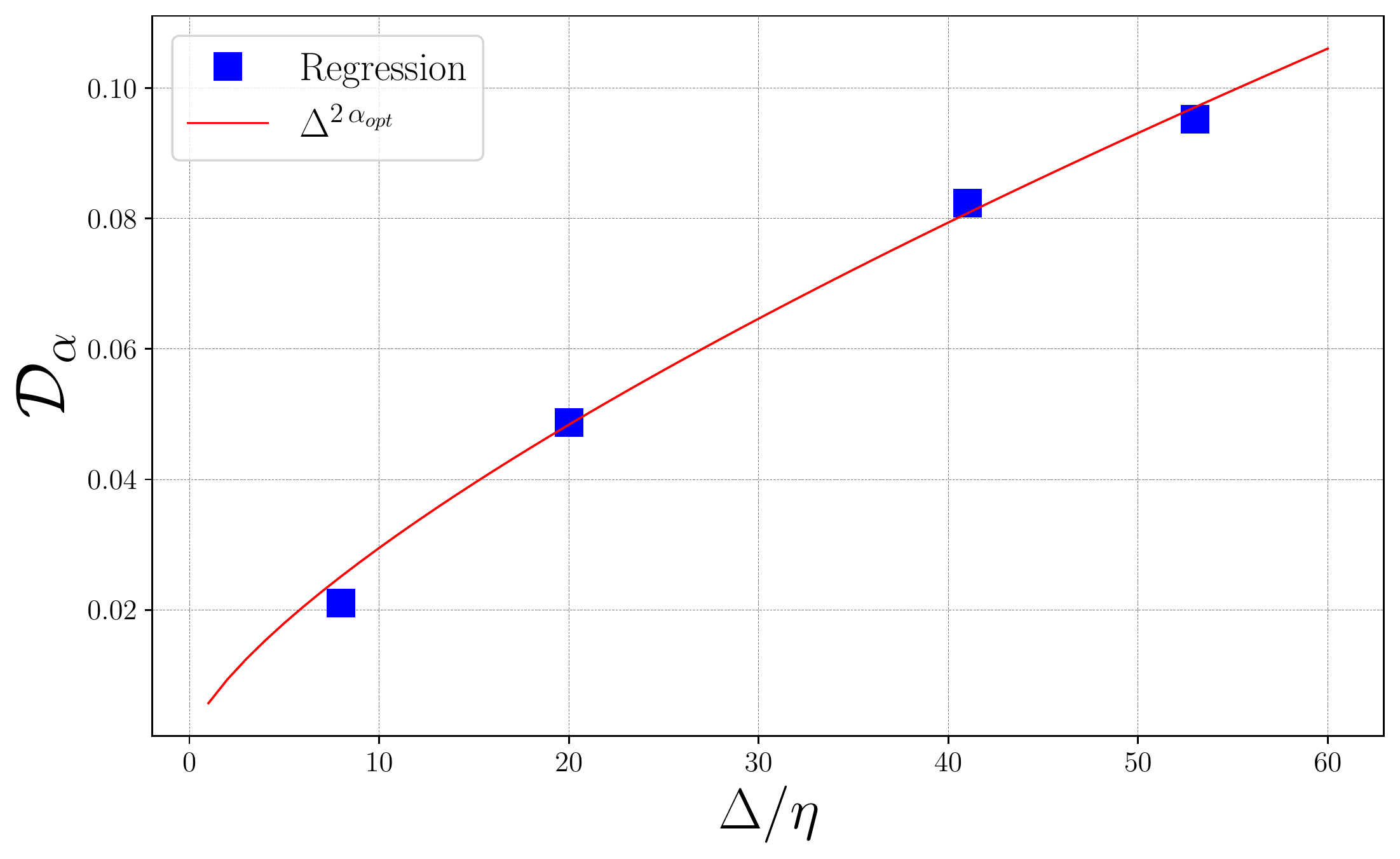}
    \caption{Variation of the proportionality coefficient, $\mathcal{D}_\alpha$, for fractional-order SGS model with filter width, and the scale invariance study}.\label{fig: D-alpha}
\end{figure}

Using the described procedure, the proportionality coefficient for each filter width is achieved. Figure \ref{fig: D-alpha} illustrates predicted $\mathcal{D}_\alpha$ through this regression procedure as a function of chosen filter width, and it is notable that the predicted $\mathcal{D}_\alpha$ decreases to lower values as we chose smaller filter widths. This numerical observation is consistent with our theoretical interpretation of $\mathcal{D}_\alpha$ as we pointed out in section \ref{sec: Derivation}. A vital consideration in developing an SGS model is the concept of scale-invariant closure model, especially within the inertial-convective subrange \citep{meneveau2000scale}. As indicated in section \ref{sec: Derivation}, $\mathcal{D}_\alpha$ takes the unit of [$L^{2\alpha}/T$]. Therefore, to study the scale invariance property, by choosing the filter width as the length-scale, one can compare the variations of $\mathcal{D}_\alpha$ obtained from the sparse regression against $\Delta^{2\alpha_{opt}}$. Figure \ref{fig: D-alpha} shows that the developed fractional-order SGS model is scale-invariant.

\section{\textit{A Priori} Testing via SGS Dissipation of the Resolved Scalar Variance}\label{sec: Dissipation}

\begin{figure}[t!]
    \begin{minipage}[b]{.49\linewidth}
        \centering
        \includegraphics[width=1\textwidth]{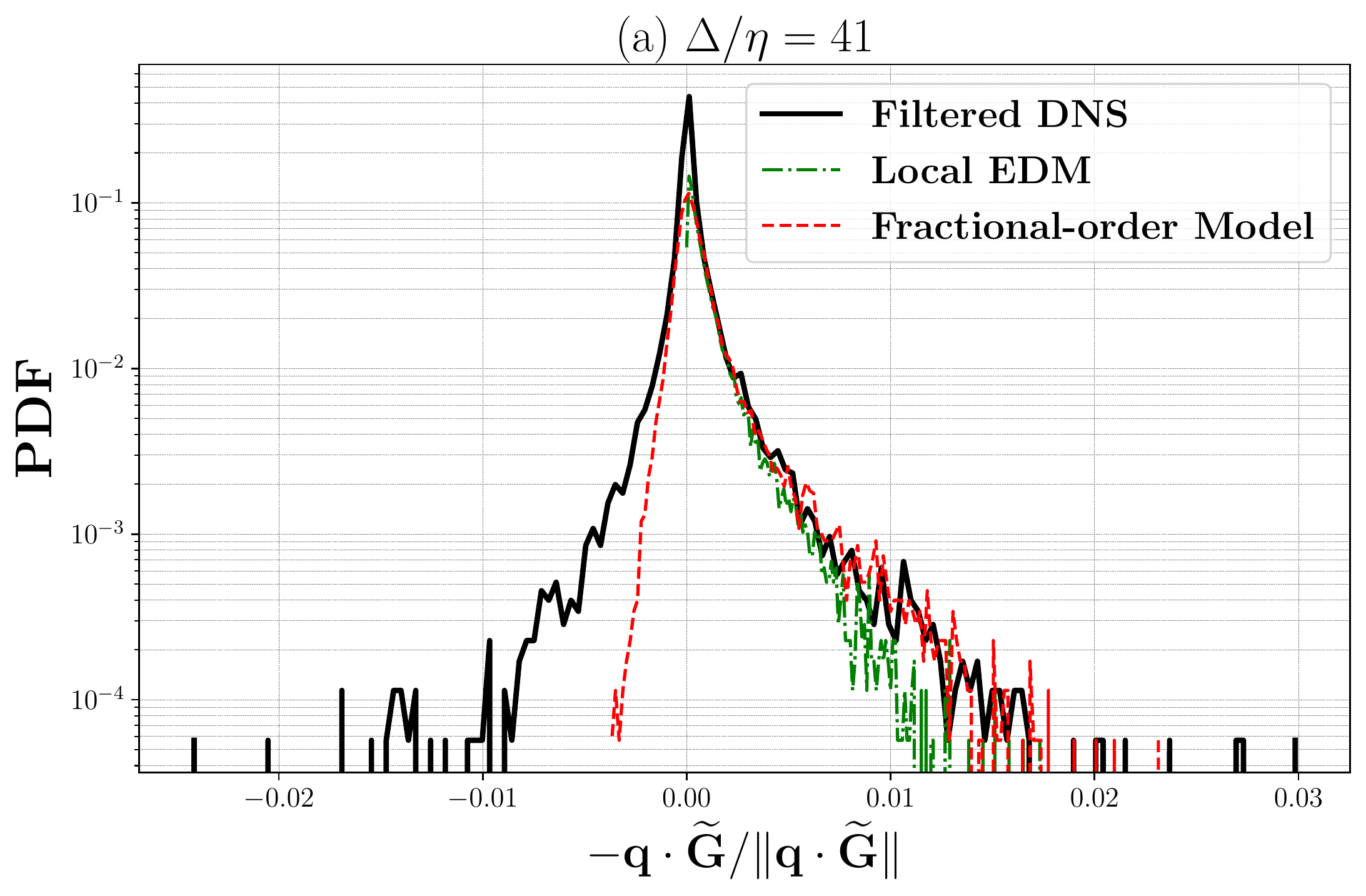}
    \end{minipage}
    \begin{minipage}[b]{.02\linewidth}
    		~
    \end{minipage}
    \begin{minipage}[b]{.49\linewidth}
        \centering
        \includegraphics[width=1\textwidth]{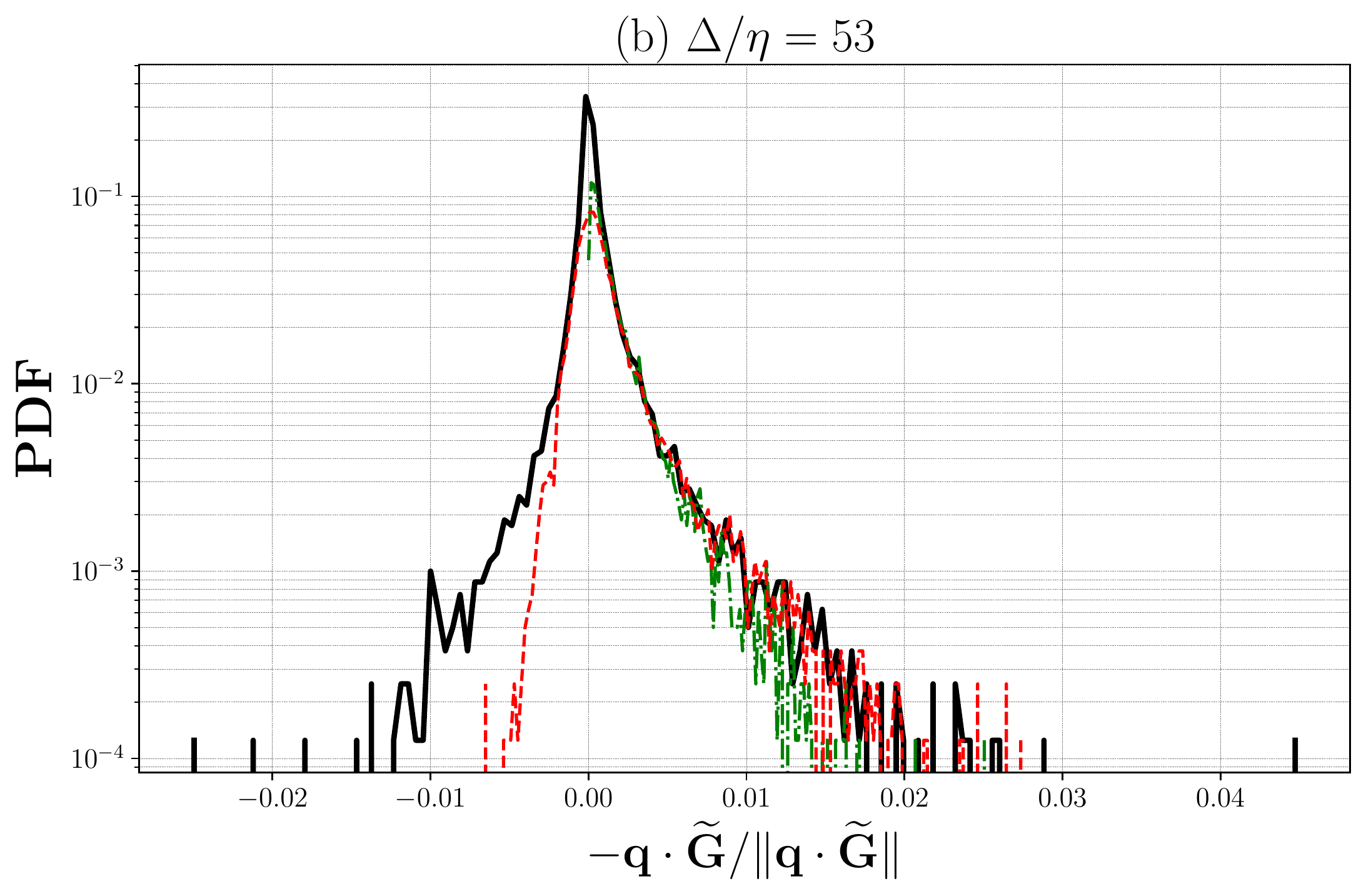}
    \end{minipage}   
    \caption{\footnotesize Probability distribution functions of the SGS dissipation of scalar variance, for the exact values from filtered DNS, local eddy-diffusivity model, and fractional-order SGS model at filter widths $\Delta/\eta=41, \, 53$.}\label{fig: Diss_PDFs}
\end{figure}

We subsequently examine the capability of the optimal fractional SGS model in reproducing the PDF of SGS dissipation of scalar variance, $\boldsymbol{q}^R \cdot \widetilde{\boldsymbol{G}}$. Through addressing:
\begin{itemize}
	\item The ability of the SGS model to capture heavy tails in the true PDF, and
	\item If the SGS model is capable of representing the backward scattering of the scalar variance cascade i.e., reproducing the negative values in the PDF.
\end{itemize}
Considering two filter widths of $\Delta/\eta=41, \, 53$, Figure \ref{fig: Diss_PDFs} shows the PDF of normalized SGS dissipation of filtered scalar variance for the optimal fractional-order model, local EDM, and the true SGS flux. The sample space to compute the PDFs is identical to the one we utilized to obtain the PDFs illustrated in Figure \ref{fig: Nonlocality}a as fully described in section \ref{sec: nonlocality}. Here, one can see that for both of the filter widths the fractional-order SGS model successfully captures the broad tail of the PDF in the positive value region for the SGS dissipation, however, the local eddy-diffusivity model fails to completely do that. The positive side of the PDF is associated with the cascade of scalar variance from the resolved scales to the unresolved ones \textit{i.e.}, forward scattering of the scalar variance. On the other hand, this figure remarkably demonstrates that unlike the local EDM, fractional-order SGS model is able to predict the events with the negative SGS dissipation values as observed in the true SGS dissipation PDFs. In fact, our resulting PDFs display that the nonlocal modeling of the SGS scalar flux through fractional-order operator makes it possible to include the backward scattering in the LES of turbulent scalar transport. Similar observation in the context of fractional-order SGS modeling was reported by Di Leoni \textit{et al.}, where their developed fractional SGS model was shown to be able to reproduce the back-scattering of the filtered turbulent kinetic energy \citep{di2020two}.

\section{Conclusions and Remarks}\label{sec: Conclusion}

We developed a new data-driven nonlocal/fractional SGS model for the LES of passive scalars transported in the homogeneous isotropic turbulent flow. The main focus of our work was on obtaining an SGS model that is structurally designed based on the nonlocal nature of the SGS scalar flux. Therefore, we first managed to present a through statistical interpretation of nonlocality in the SGS dynamics using the single- and two-point statistics of the SGS scalar dissipation. Using a rich dataset of high-fidelity data for the SGS flux obtained from direct filtering of DNS results, we illustrated the statistical nonlocality embedded in the SGS dynamics and showed that it amplifies as the filter-width increases. Moreover, we showed that the conventional means of SGS modeling originate from a local statistical representation for the SGS dynamics and are intrinsically incapable of predicting the statistical nonlocality. As a robust starting point for our mathematical modeling, we started from Boltzmann-BGK kinetics as the microscopic transport framework for passive scalars in homogeneous turbulence and considered the closure problem manifested in filtering the transport equations. By revisiting the kinetic-level strategy for the LES modeling taking into account the consistency of the the model for the filtered equilibrium distribution with its macroscopic representation at the continuum level, we proposed to proceed with closure modeling using $\alpha$-stable L\'evy distribution to address the nonlocal and non-Gaussian behavior of the closure at the kinetic level. In order to derive a macroscopic representation of such model to employ in the filtered AD equation, we used continuum averaging and obtained the filtered and residual (modeled) passive scalar flux components that essentially return the filtered AD equation. Throughout this procedure, the up-scaled model for the divergence of the residual flux takes the form of a fractional Laplacian acting on the filtered scalar concentration with a model-specific proportionality coefficient. Next, we managed to calibrate the fractional-order model in two separate data-driven stages. First we targeted identification of the optimal fractional order using two-point statistics data for the normalized SGS dissipation function obtained from the DNS and minimizing the mismatch function with its counterpart in the fractional-order SGS model. This procedure returned the optimal fractional order that minimizes the single-point correlation between the modeled and true SGS scalar flux. Afterwards, following an sparse regression strategy over the spatio-temporal data for the SGS scalar flux in a statistically-stationary turbulent scalar field, we obtained the proportionality coefficient of the model. Moreover, we showed the consistency of the derived model in terms of the relationship between the obtained proportionality coefficient and decreasing the filter-width. Finally, in an \textit{a priori} test, we showed that the identified model is capable of capturing the PDF tail associated with the forward scattering of the filtered scalar variance and illustrated that our model has the capability to partially reproduce the backward scattering phenomenon.

\section*{Acknowledgement}

This work was financially supported by the MURI/ARO grant (W911NF-15-1-0562), the ARO Young Investigator Program (YIP) award (W911NF-19-1-0444), and partially by the National Science Foundation award (DMS-1923201). The high-performance computing resources and services were provided by the Institute for Cyber-Enabled Research (ICER) at Michigan State University.

\appendix

\section{Fractional-Order Differential Operators} \label{sec: Fractional-Calc}

According to Lischke \textit{et al.} \citep{lischke2018fractional}, the fractional Laplacian operator, denoted by $(-\Delta)^{\alpha}$ with $0 < \alpha \leq 1$, is defined as
\begin{eqnarray}\label{FL-1}
    (-\Delta)^{\alpha} u(\boldsymbol{x}) &=& \frac{1}{(2 \pi)^d} \int_{\mathbb{R}^d } \vert \boldsymbol{\xi} \vert^{2\alpha} \, \big {(} u,\, e^{-\mathfrak{i} \boldsymbol{\xi}\cdot \boldsymbol{x} } \big {)} \, e^{\mathfrak{i} \boldsymbol{\xi}\cdot \boldsymbol{x}} \,  d\boldsymbol{\xi}
    \nonumber \\
    &=& \mathcal{F}^{-1} \Big {\{}   \vert \boldsymbol{\xi} \vert^{2\alpha} \mathcal{F}  \big {\{}u\big {\}} (\boldsymbol{\xi}) \Big {\}},
\end{eqnarray}
where $\mathcal{F}$ and $\mathcal{F}^{-1}$ represent the Fourier and inverse Fourier transforms for a real-valued vector $\boldsymbol{\xi}=\xi_j$, $j=1,\, 2,\, 3$, respectively, and $\mathfrak{i}=\sqrt{-1}$. Moreover, $ ( \cdot \, , \cdot )$ specifies the $L_2$-inner product on $\mathbb{R}^d$, $d=1,2,3$. Therefore, the Fourier transform of the fractional Laplacian is then obtained as
\begin{equation}\label{FL-2}
    \mathcal{F} \Big {\{} (-\Delta)^{\alpha} u(\boldsymbol{x})  \Big {\}}=\vert \boldsymbol{\xi} \vert^{2\alpha} \mathcal{F}  \big {\{}u\big {\}} (\boldsymbol{\xi}),
\end{equation}
where $\alpha=1$ recovers the integer-order Laplacian. Considering the definition of $\alpha$-Riesz potential as
\begin{eqnarray}\label{FL-3-2}
    \mathcal{I}_{\alpha} u(\boldsymbol{x}) &=& C_{d,-\alpha} \, \int_{\mathbb{R}^d }\frac{u(\boldsymbol{x})-u(\boldsymbol{s})}{\vert \boldsymbol{x}-\boldsymbol{s}\vert^{d-2\alpha}} ds,
\end{eqnarray}
the fractional Laplacian can also be expressed in the integral form as 
\begin{eqnarray}\label{FL-3}
    (-\Delta)^{\alpha} u(\boldsymbol{x}) &=& C_{d,\alpha} \, \int_{\mathbb{R}^d }\frac{u(\boldsymbol{x})-u(\boldsymbol{s})}{\vert \boldsymbol{x}-\boldsymbol{s}\vert^{2\alpha+d}} d\boldsymbol{s},
\end{eqnarray}
where $C_{d,\alpha} = \frac{2^{2\alpha} \Gamma(\alpha+d/2)}{\pi^{d/2} \Gamma(-\alpha)}$ for $\alpha \in (0,1]$ and $\Gamma(\cdot)$ represents Gamma function \citep{lischke2018fractional}. The $\alpha$-Riesz potential is also formulated \citep{stein2016singular} as
\begin{eqnarray}\label{FL-3-3}
    \mathcal{I}_{\alpha} u(\boldsymbol{x}) &=& (-\Delta)^{-\alpha} u(\boldsymbol{x}) = \mathcal{F}^{-1} \Big {\{}   \vert \boldsymbol{\xi} \vert^{-2\alpha} \mathcal{F}  \big {\{}u\big {\}} (\boldsymbol{\xi}) \Big {\}}.
\end{eqnarray}
Considering \eqref{FL-3-3}, the Riesz transform is then given by
\begin{eqnarray}\label{FL-3-4}
    \mathcal{R}_j u(\boldsymbol{x}) &=&\nabla_j \, \mathcal{I}_{1} u(\boldsymbol{x}) = \mathcal{F}^{-1} \Big {\{}  -\frac{\mathfrak{i}\xi_j}{ \vert \boldsymbol{\xi} \vert} \mathcal{F}  \big {\{}u\big {\}} (\boldsymbol{\xi}) \Big {\}},
\end{eqnarray}
which is utilized in formulating the SGS scalar flux.

\section{Derivation of Passive Scalar Flux}\label{sec: Appendix1}

Regarding the filtered passive scalar flux given in \eqref{GE-25}, one can write that
\begin{eqnarray}\label{GE-27}
    \widetilde{q_i} &=& \int_{0}^{\infty} \int_{\mathbb{R}^d} (u_i-\widetilde{V}_i) \left(\widetilde{\Phi_{s,s}} \, F(\widetilde{\mathcal{L}_{s,s}})-\widetilde{\Phi} \, F(\widetilde{\mathcal{L}})\right) \, e^{-s} d\boldsymbol{u} \, ds,
\end{eqnarray}
where using the Taylor expansions of $\widetilde{\Phi_{s,s}}$ and $F(\widetilde{\mathcal{L}_{s,s}})$ about their not shifted values and later on by utilizing the incompressibility constraint, one arrives at the following
\begin{eqnarray}\label{GE-28}
    \widetilde{q_i} &=& \int_{0}^{\infty}  \int_{\mathbb{R}^d} (u_i-\widetilde{V}_i) \left(-u_j \, s\tau_g\right) \, \frac{\partial \widetilde{\Phi}}{\partial x_j} \, F(\widetilde{\mathcal{L}}) \, e^{-s} \, d\boldsymbol{u}\, ds
    \\ \nonumber
    &\simeq& -\frac{\tau_g}{c_T^3}\frac{\partial \widetilde{\Phi}}{\partial x_i} \left(\int_{0}^{\infty} s \, e^{-s}\, ds\right) \int_{\mathbb{R}^d} (u_j-\widetilde{V}_j)(u_j-\widetilde{V}_j) \, F(\widetilde{\mathcal{L}}) \, d\boldsymbol{u}.
\end{eqnarray}
Knowing that $\int_{0}^{\infty} s \, e^{-s}\, ds=1$, the \textit{diffusivity} coefficient of the passive scalar, $\mathcal{D}$, would be expressed as
\begin{align}\label{GE-29}
    \mathcal{D} := \frac{\tau_g}{c_T^3} \, \int_{\mathbb{R}^d} (u_j-\widetilde{V}_j)(u_j-\widetilde{V}_j) \, F(\widetilde{\mathcal{L}}) \, d\boldsymbol{u}.
\end{align}
As a result, the filtered (resolved) passive scalar flux, $\widetilde{\boldsymbol{q}}$, and its divergence appearing in the right-hand side of \eqref{GE-17} could be written as
\begin{align}
    \widetilde{q}_i = -\mathcal{D} \, \frac{\partial \widetilde{\Phi}}{\partial x_i} \quad \Longrightarrow \quad \nabla \cdot \widetilde{\boldsymbol{q}} = -\mathcal{D} \, \Delta \, \widetilde{\Phi}.
\end{align}
On the other hand, the integral form of the modeled SGS flux in \eqref{GE-26} can be written in the following form
\begin{eqnarray}\label{GE-30}
    q_i^R &=& \int_{0}^{\infty}  \int_{\mathbb{R}^d} (u_i-\widetilde{V}_i) \left(\widetilde{\Phi_{s,s}} \, F^\alpha(\widetilde{\mathcal{L}_{s,s}})-\widetilde{\Phi} \, F^\alpha(\widetilde{\mathcal{L}})\right) e^{-s} d\boldsymbol{u}\, ds,
\end{eqnarray}
By adding and subtracting $\widetilde{\Phi_{s,s}} \, F^\alpha(\widetilde{\mathcal{L}})$ to $\left(\widetilde{\Phi_{s,s}} \, F^\alpha(\widetilde{\mathcal{L}_{s,s}})-\widetilde{\Phi} \, F^\alpha(\widetilde{\mathcal{L}})\right)$, one can rewrite \eqref{GE-30} as
\begin{eqnarray}\label{GE-31}
    q_i^R &=& \int_{0}^{\infty}  \int_{\mathbb{R}^d} (u_i-\widetilde{V}_i) \left(\widetilde{\Phi_{s,s}}-\widetilde{\Phi}\right)F^\alpha(\widetilde{\mathcal{L}}) \, e^{-s} d\boldsymbol{u}\, ds
    \\ \nonumber
    && + \int_{0}^{\infty}  \int_{\mathbb{R}^d} (u_i-\widetilde{V}_i) \, \widetilde{\Phi_{s,s}} \left(F^\alpha(\widetilde{\mathcal{L}_{s,s}})-F^\alpha(\widetilde{\mathcal{L}})\right) \, e^{-s} d\boldsymbol{u}\, ds,
\end{eqnarray}
where the second integral is approximated as zero. Therefore, the modeled subgrid-scale scalar flux is simplified into
\begin{eqnarray}\label{GE-33}
    q_i^R &=& \int_{0}^{\infty}  \int_{\mathbb{R}^d} (u_i-\widetilde{V}_i) \left(\widetilde{\Phi_{s,s}}-\widetilde{\Phi}\right) \, F^\alpha(\widetilde{\mathcal{L}}) \, e^{-s} d\boldsymbol{u}\, ds.
\end{eqnarray}
Considering $u_i=(x_i^\prime-x_i)/s\tau$ and approximating $u_i-\widetilde{V_i}\simeq u_i$, one can obtain that $d\boldsymbol{u}=d\boldsymbol{x}^\prime/(s\tau_g)^3$ \citep{epps2018turbulence, samiee2020fractional}. As a result, $\widetilde{\mathcal{L}}=( \boldsymbol{u}-\widetilde{V})^2/c_T^2 \approx \boldsymbol{u}^2/c_T^2 = ( \boldsymbol{x}^\prime-\boldsymbol{x})^2/(s \, \tau_g \, c_T)^2$. According to the definition of isotropic $\alpha$-stable L\'evy distribution, $F^\alpha(\widetilde{\mathcal{L}})=C_\alpha/\widetilde{\mathcal{L}}^{(2\alpha+3)/2}$, where $C_\alpha$ is a real-valued constant. Consequently, \eqref{GE-33} may be reformulated as
\begin{eqnarray}\label{GE-34}
    q_i^R &=& \int_{0}^{\infty}  \int_{\mathbb{R}^d} \frac{1}{c_T^3 \, s^3 \, \tau_g^3} \left(\frac{x_i^\prime-x_i}{s \, \tau_g}\right) \left(\widetilde{\Phi_{s,s}}-\widetilde{\Phi}\right) \left( \frac{C_\alpha}{\widetilde{\mathcal{L}}^{(2\alpha+3)/2}}\right) e^{-s} d\boldsymbol{x}^\prime \, ds
    \\ \nonumber
    &=& -\frac{C_\alpha (c_T \, \tau_g)^{2\alpha}}{\tau_g} \left(\int_{0}^{\infty} e^{-s}s^{2\alpha-1} ds \right)     \int_{\mathbb{R}^d} \frac{(x_i^\prime-x_i) \left(\widetilde{\Phi}(\boldsymbol{x}^\prime)-\widetilde{\Phi}(\boldsymbol{x})\right)}{\vert \boldsymbol{x}^\prime-\boldsymbol{x} \vert^{2\alpha+3}} \, d\boldsymbol{x}^\prime
    \\ \nonumber
    &=& -\frac{C_\alpha (c_T \, \tau_g)^{2\alpha}}{\tau_g} \, \Gamma(2\alpha) \int_{\mathbb{R}^d} \frac{(x_i^\prime-x_i) \left(\widetilde{\Phi}(\boldsymbol{x}^\prime)-\widetilde{\Phi}(\boldsymbol{x})\right)}{\vert \boldsymbol{x}^\prime-\boldsymbol{x}\vert^{2\alpha+3}} \, d\boldsymbol{x}^\prime,
\end{eqnarray}
By taking the divergence of the modeled SGS scalar flux in \eqref{GE-34}, we obtain
\begin{eqnarray}\label{GE-35}
    (\nabla \cdot \boldsymbol{q}^R)_i &=& -\frac{C_\alpha (c_T \, \tau_g)^{2\alpha}}{\tau_g} \, \Gamma(2\alpha) \, \nabla \cdot \int_{\mathbb{R}^d} \frac{(x_i^\prime-x_i) \left(\widetilde{\Phi}(\boldsymbol{x}^\prime)-\widetilde{\Phi}(\boldsymbol{x})\right)}{\vert \boldsymbol{x}^\prime-\boldsymbol{x}\vert^{2\alpha+3}} \, d\boldsymbol{x}^\prime
    \\ \nonumber
    &=& -\frac{C_\alpha (c_T \, \tau_g)^{2\alpha}}{\tau_g} \, \Gamma(2\alpha) \, \left[ (2\alpha+2)\int_{\mathbb{R}^d} \frac{\widetilde{\Phi}(\boldsymbol{x}^\prime)-\widetilde{\Phi}(\boldsymbol{x})}{\vert \boldsymbol{x}^\prime-\boldsymbol{x}\vert^{2\alpha+3}} d\boldsymbol{x}^\prime -\int_{\mathbb{R}^d} \frac{\partial \widetilde{\Phi}/\partial x_i}{\vert \boldsymbol{x}^\prime-\boldsymbol{x}\vert^{2\alpha+2}} \,  d\boldsymbol{x}^\prime \right].
\end{eqnarray}
Due to symmetry, the second integral inside the large brace is zero and the other integral is nothing but the definition of fractional Laplacian of the filtered passive scalar concentration, $\widetilde{\Phi}$. Thus, \eqref{GE-35} takes the following compact form
\begin{align}\label{GE-36}
    \nabla \cdot \boldsymbol{q}^R = -\frac{C_\alpha (c_T \, \tau_g)^{2\alpha}}{2\tau_g} (2\alpha+2)\, \Gamma(2\alpha) \, (-\Delta)^{\alpha} \widetilde{\Phi}, \quad \alpha \in (0,1].
\end{align}


\bibliographystyle{elsarticle-num}
\bibliography{mybibfile}

\end{document}

%% file: preamble.tex
\usepackage{lineno}
\modulolinenumbers[5]
\usepackage{epstopdf}
\usepackage{subcaption}
\usepackage{tabularx}
\usepackage{graphics} 
\usepackage{graphicx} 
\usepackage{caption}
\usepackage{subcaption}
\usepackage{amsmath,amssymb,amstext,amsthm,mathtools}
\usepackage{float}
\usepackage{booktabs}
\usepackage{multirow}
\usepackage{siunitx} 
\usepackage{array,bm,color,epsfig}
\usepackage{caption}
\usepackage{url}
\usepackage[colorlinks]{hyperref}
\usepackage[english]{babel}
\usepackage[utf8x]{inputenc}
\usepackage{setspace}
\usepackage{filecontents}
\usepackage{natbib}
\usepackage{float}
\usepackage{listings}
\usepackage{color}
\usepackage{xcolor}
\usepackage{wasysym}
\usepackage[flushleft]{threeparttable}
\usepackage{hhline}
\usepackage{multicol,tabularx,capt-of,ragged2e,ar}
\usepackage{mathrsfs}
\usepackage{tikz}
\usepackage{pifont}
\usepackage{stmaryrd}
\usepackage{pdflscape}
\usepackage{mathrsfs}
\usepackage{stmaryrd}
\usepackage{fancyheadings}
\usepackage{wrapfig,lipsum,booktabs}
\usepackage{soul}
\usepackage[toc]{appendix}

\usepackage{scalerel}
\usepackage{stackengine,wasysym}
\newcommand\reallywidetilde[1]{\ThisStyle{%
  \setbox0=\hbox{$\SavedStyle#1$}%
  \stackengine{-.1\LMpt}{$\SavedStyle#1$}{%
    \stretchto{\scaleto{\SavedStyle\mkern.2mu\AC}{.5150\wd0}}{.6\ht0}%
  }{O}{c}{F}{T}{S}%
}}

\def\BE{\begin{equation}}
\def\EE{\end{equation}}

\theoremstyle{remb}
\newtheorem{remb}{Remark}

\theoremstyle{assump}

\theoremstyle{definition}